\documentclass[journal]{IEEEtran}
\usepackage{cite}
\usepackage{amsmath,amssymb,amsfonts}
\usepackage{algorithmic}
\usepackage{graphicx}
\usepackage{textcomp}
\usepackage{xcolor}
\usepackage{subfigure}
\usepackage{multirow}
\usepackage{stfloats}
\usepackage{siunitx}
\usepackage{float}
\usepackage{verbatim}
\usepackage{dutchcal}
\usepackage{fancyhdr}
\begin{document}

\title{Prototyping and Experimental Results for ISAC-based Channel Knowledge Map}

\author{Chaoyue~Zhang,     
Zhiwen~Zhou,       Xiaoli~Xu,~\IEEEmembership{Member,~IEEE,}\\ 
Yong~Zeng,~\IEEEmembership{Fellow,~IEEE,}
Zaichen~Zhang,~\IEEEmembership{Senior~Member,~IEEE,}
and~Shi~Jin,~\IEEEmembership{Fellow,~IEEE}

\thanks{ 
This work was supported by the Natural Science Foundation for Distinguished Young Scholars of Jiangsu Province with grant number BK20240070.
(Corresponding author: Xiaoli Xu.)

Chaoyue~Zhang, Zhiwen~Zhou, Xiaoli~Xu, Yong~Zeng, Zaichen~Zhang and Shi~Jin are with the National Mobile Communications Research Laboratory, Southeast University, Nanjing 210096, China (e-mail: \{chaoyue\underline{~}zhang, zhiwen\underline{~}zhou, xiaolixu, yong\underline{~}zeng, zczhang, jinshi\}@seu.edu.cn).

Yong~Zeng and Zaichen~Zhang are also with the Purple Mountain Laboratories, Nanjing 211111, China.}
}

 



\maketitle

\thispagestyle{fancy}

\cfoot{Copyright \textcircled c 2025 IEEE. Personal use of this material is permitted. However, permission to use this material for any other purposes must be obtained from the IEEE by sending a request to pubs-permissions@ieee.org.}

\renewcommand{\headrulewidth}{0mm}

\begin{abstract}
Channel knowledge map (CKM) is a novel approach for achieving environment-aware communication and sensing. 
This paper presents an integrated sensing and communication (ISAC)-based CKM prototype system, demonstrating the mutualistic relationship between ISAC and CKM. 
The system consists of an ISAC base station (BS), a user equipment (UE), and a server. 
By using a shared orthogonal frequency division multiplexing (OFDM) waveform over the millimeter wave (mmWave) band, the ISAC BS is able to communicate with the UE while simultaneously sensing the environment and acquiring the UE's location. 
The prototype showcases the complete process of the construction and application of the ISAC-based CKM.
For CKM construction phase, the BS stores the UE's channel feedback information in a database indexed by the UE's location, including beam indices and channel gain.  
For CKM application phase, the BS looks up the best beam index from the CKM based on the UE's location to achieve training-free mmWave beam alignment. 
The experimental results show that ISAC can be used to construct or update CKM while communicating with UEs, and the pre-learned CKM can assist ISAC for training-free beam alignment.

\end{abstract}


%

\section{Introduction}

\begin{figure*}[!hb]
	\centering
    \label{ckm_apply}
    \subfigure[ISAC-based CKM construction.]{
        \label{ckm_apply_los}
        \includegraphics[width=0.30\textwidth]{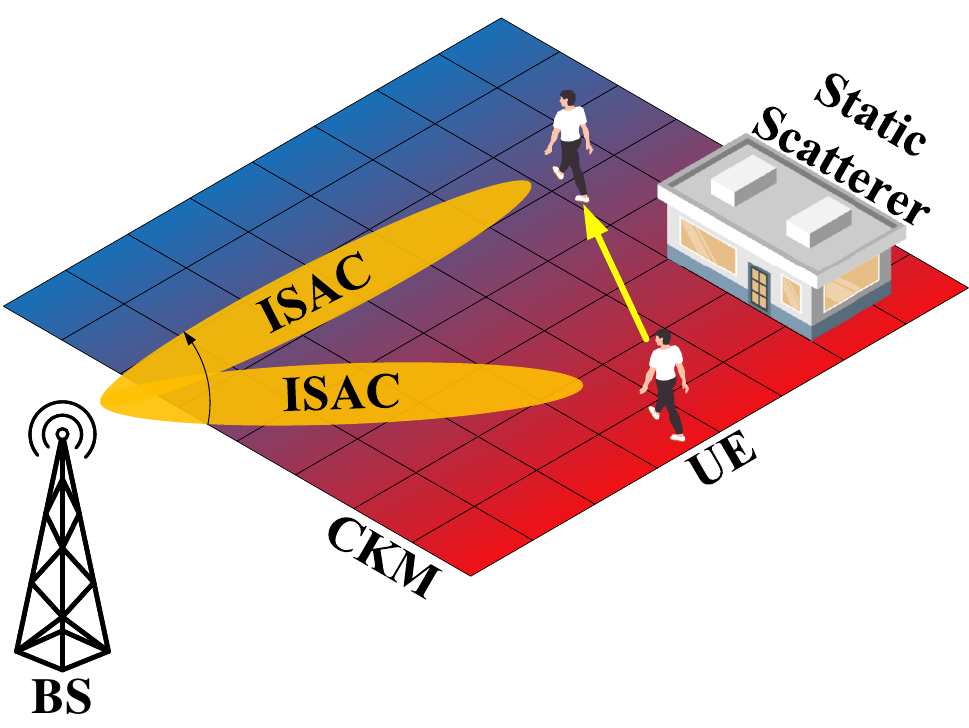}
		}
	\hfill
	\subfigure[CKM application in ISAC system with occasionally blocked LoS link.]{
    \label{ckm_apply_dyna}
    \includegraphics[width=0.30\textwidth]{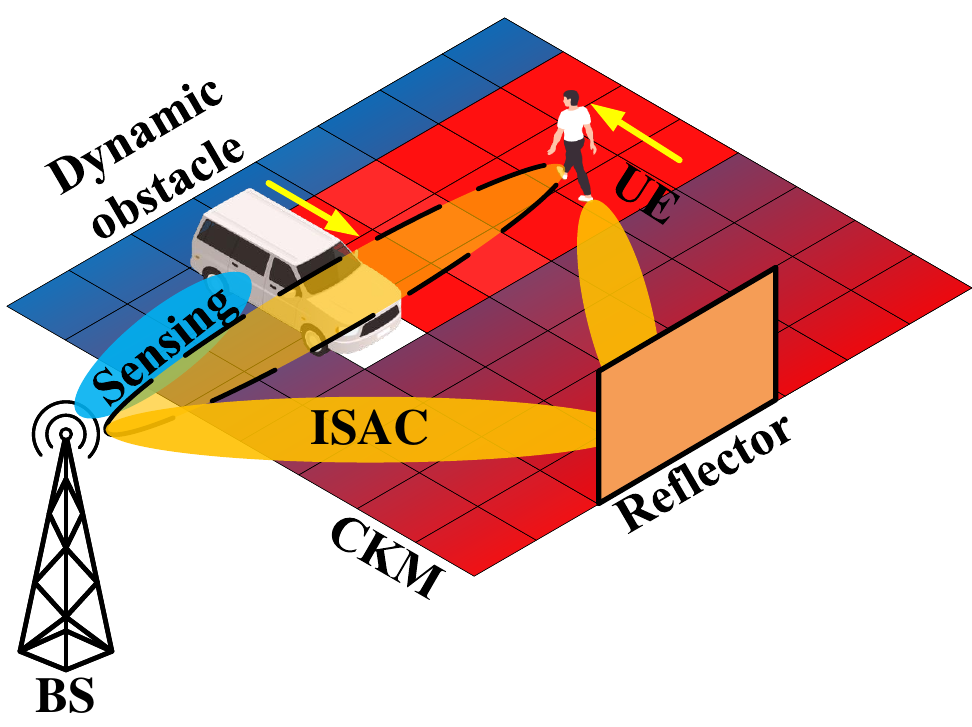} 
	}
    \hfill
    \subfigure[CKM application in ISAC system with NLoS link.]{
    \label{ckm_apply_nlos}
    \includegraphics[width=0.30\textwidth]{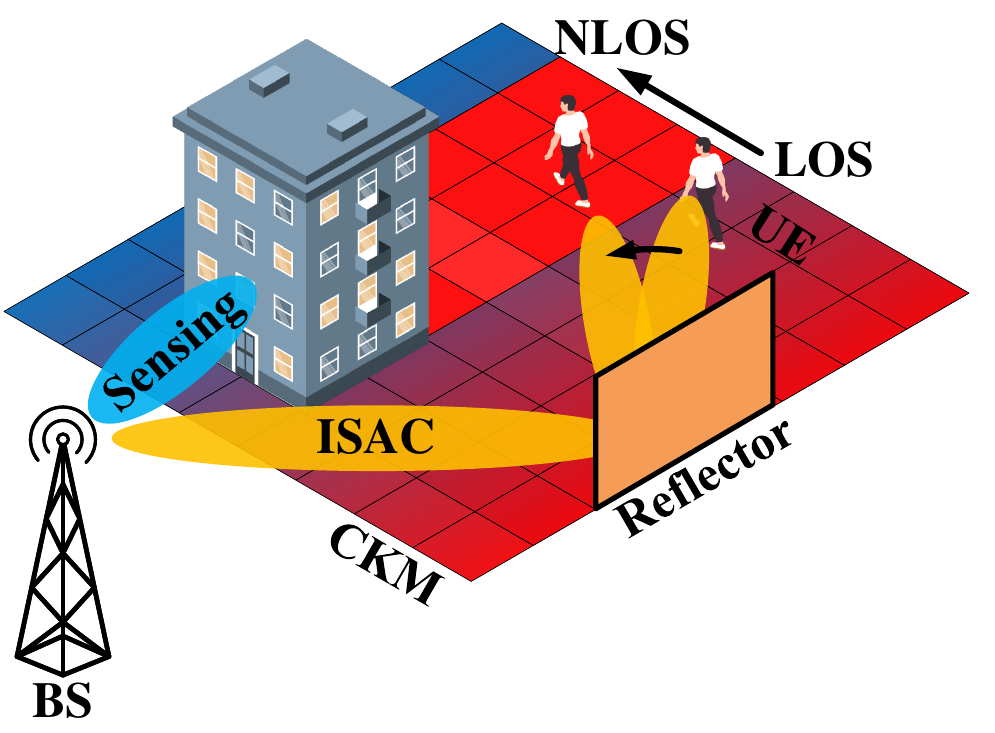}}
    \caption{Typical scenarios for ISAC-based CKM prototyping.}
    
\end{figure*}

Channel knowledge map (CKM) is a site-specific database, tagged with the locations of the transmitters and/or receivers, that contains channel-related information useful to enhance environment-awareness and facilitate or even eliminate the process of sophisticated real-time channel state information (CSI) acquisition\cite{CKM}\cite{ckm_tutorial}.
With CKM, the environment-aware channel knowledge can be inferred directly based on physical or virtual location information of the mobile terminals. 
Typical channel knowledge includes channel gain, path loss, angle-of-arrival (AoA)/angle-of-departure (AoD), beam indices.
For example, for millimeter wave (mmWave) massive  multiple input multiple output (MIMO) systems, with CKM, base station (BS) and user equipment (UE) can obtain candidate beam indices based on their locations and implement refined beam sweeping over a much smaller beam set \cite{CKM_beamforming}, instead of exhaustively searching all the beam pairs. 
Compared to conventional radio environment maps (REMs) that mainly concern spectrum usage information \cite{Engineering_Radio_Maps}\cite{REM}, CKM focuses on the location-specific wireless channel knowledge that is independent of the transmitter/receiver activities. 

Extensive efforts have been devoted to the theoretical studies of CKM construction and utilization. 
For example, \cite{how_much_ckm} investigates the amount of data required to construct the particular type of CKM, namely channel gain map (CGM).
The work \cite{CKM_beam_muti_alignment} proposes an environment-aware coordinated multi-point (CoMP) beam alignment scheme for mmWave systems, by leveraging the novel technique of beam index map (BIM). 
\cite{Video_ckm} applies CKM to the scenario of aerial video streaming. 
In \cite{ris_ckm}, the authors propose a novel environment-aware joint active/passive beamforming for RIS-aided wireless communication based on CKM. 
In our previous work \cite{exp_ckm}, we have developed a preliminary prototype system to construct BIM for mmWave communication systems to learn the candidate transmit and receive beam index pairs for each grid location in the experiment area.
However, existing CKM construction methods mainly rely on external localization systems such as ultra-wide band (UWB) systems to acquire the location of the mobile terminals, which means that the terminals need to be cooperative and they have to carry localization tags. 
Besides, such systems are unable to achieve sensing and localization of environmental scatterers. 
Therefore, more effective approaches for localization and environment sensing are need for the widespread application of CKM.

In the meantime, integrated sensing and communication (ISAC) has been identified as one of the main usage scenarios for the sixth-generation (6G) mobile communication networks \cite{IMT2030}.
One of the main objectives of ISAC system is to estimate various parameters of the sensing targets based on echo signals, such as the target distance, velocity, angle, and radar cross section (RCS) \cite{ISAC_in_6g}. 
The `target' includes not only the UE but also any scatterer in the environment.
Such parameters can provide valuable prior information for communication and enables efficient channel estimation and beam alignment. 
According to different goals of communication and sensing, ISAC can be distinguished into two specific application scenarios: the ISAC beam tracking scenario for users, and the ISAC beam scanning scenario for the surrounding environment. 
To establish the initial access for UEs, ISAC BS needs to perform beam sweeping in the environment to locate the UE and image the surroundings. 
Once the user has established the initial access, the BS switches to beam tracking towards UE while maintaining communication with it and sensing its location. 
To maintain awareness of the dynamic environment, the BS still performs beam sweeping at regular intervals. 
ISAC BS transmits data while sensing the environment, provides UE's position and environment information.
We consider several typical scenarios where a mmWave ISAC BS serves a mobile UE.
The primary objective of the BS is to ensure reliable communication services for the UE by maintaining continuous beam alignment. 
For CKM construction shown in \mbox{Fig.~\ref{ckm_apply_los}}, the UE's location and channel feedback are needed. 
ISAC can be used to efficiently obtain the UE's location, and the channel training resource can be used for environment sensing.
Hence the ISAC BS can build the CKM while communicating with the mobile UE.
For CKM application in ISAC system, UE's location and environment information can be used for communication and sensing beam tracking of UE. 
As shown in \mbox{Fig.~\ref{ckm_apply_dyna}}, the LoS link between ISAC BS and UE may be occasionally blocked by dynamic obstacles.
CKM enables proactive beam switching while UE enters non-line-of-sight (NLoS) area without having to scan the entire environment again. 
As shown in \mbox{Fig.~\ref{ckm_apply_nlos}}, if the UE moves to the NLoS area, ISAC BS can directly obtain the beam corresponding to the strongest NLoS path based on the location information provided by the UE, which enables training-free beam alignment and continuous ISAC services. 

\begin{figure}[htbp]
\centering
	{\includegraphics[width=0.48\textwidth]{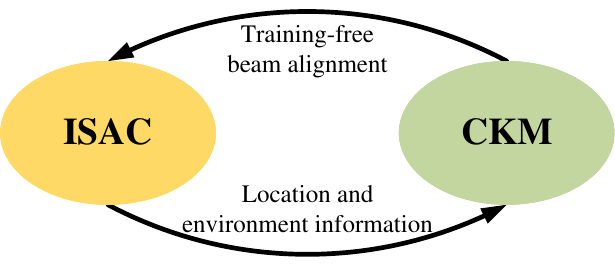}}
	\caption{ISAC based CKM application.}
	\label{fig_isac_ckm}
\end{figure}

The relationship between CKM and ISAC is inherently mutualistic, as illustrated in \mbox{Fig.~\ref{fig_isac_ckm}}. 
The BS requires the UE's location to query channel knowledge from CKM, and ISAC can be leveraged to efficiently obtain the UE's location. 
Moreover, the channel information obtained from CKM can also enhance ISAC's performance, such as training-free beam alignment and channel estimation.

There are some preliminary prototyping and experimental results reported for ISAC. 
For example, in our previous work \cite{ISSAC}, the concept of integrated super-resolution sensing and communication (ISSAC) is proposed, and the feasibility of applying super-resolution algorithm to ISAC based on new radio (NR) waveform is verified through experiments. 
Besides, \cite{An_Experimental_Study} shows the tradeoff between communication and sensing performance in ISAC system through experiments.
\cite{ISAC_MIMO_TENSOR,Channel_Training_THZ,Movable_Antenna_MIMO}  investigate the system models of massive MIMO-ISAC, terahertz massive MIMO-ISAC, and movable antenna MIMO, respectively. 
By leveraging tensor-based approaches, they significantly enhance the channel estimation accuracy in these wireless communication systems while reducing training overhead.
\cite{Tensor_Decomposition} studies the channel estimation of time-varying and frequency-selective (TVFS) mmWave MIMO channels in high-mobility scenarios with severe Doppler effects, and further proposes a joint AoD and Doppler shift estimation (JADE) algorithm with higher estimation accuracy.
\cite{First_Demonstration} introduces the MIMO radar based on orthogonal frequency division multiplexing (OFDM) waveform for high-resolution synthetic aperture radar (SAR) imaging and data transmission.
\cite{Joint_Communication} and \cite{Multifunctional_Transceiver} developed the time-division based ISAC system, and \cite{Spatial_Modulation} is frequency-division based ISAC system.
These prototypes experimentally verified the performance improvements brought by ISAC for both communication and sensing.
Recently, we presented a demo for a prototype of real-time ISAC with mmWave OFDM in \cite{exp_isac}. 
The prototype is designed to demonstrate ISAC for internet of vehicles (IoV), which achieves high-quality communication to vehicles while sensing their positions. 
Additionally, the platform is capable of real-time sensing of obstacles in the environment, thereby enhancing traffic safety in IoV scenarios. 
On the other hand, we presented a prototype system for environment-aware communication based on CKM, in \cite{exp_ckm}. 
By acquiring information about the locations and angles of UEs and major obstacles, we utilized CKM to achieve efficient mmWave beam alignment in different scenarios without real-time channel training. 
This validated the effectiveness of mmWave beam alignment based on CKM in both quasi-static and dynamic environments. 

Based on the foundations of previous ISAC and CKM prototypes, we further proposed an ISAC-based CKM prototype.\footnote{The video introduction of the developed prototype and experimental results can be found at https://youtu.be/nnOog2kKQO0.} 
This prototype demonstrates real-time construction and application of ISAC-based CKM.
The presence of CKM enables ISAC BS to achieve uninterrupted mmWave communication to UE when LoS path is blocked by dynamic obstacles. 
Further more, when UE is in NLoS area, by periodically reporting its location to ISAC BS, UE can also obtain high-performance communication services without frequent beam training.
Meanwhile, the presence of ISAC enables direct acquisition of  UE's location, eliminating the dependence on external positioning systems such as UWB for UE positioning as in \cite{exp_ckm}. 
Moreover, this prototype achieves real time construction of CKM by efficiently reusing the environment sensing signals for UE beam training.
The main contributions of this work are summarized as follows:
\begin{itemize}
    \item First, the principle of the ISAC-based CKM system is explained. 
    The communication channel, sensing channel and signal model of the ISAC system are presented, and a specific ISAC waveform for phased antenna array with one radio frequency (RF) chain is proposed. 
    Specifically, to achieve real-time environment sensing while providing continuous ISAC services to users, it is necessary to separate the processes for environment sensing and ISAC for user in the time domain. 
    Therefore, a dedicated environment sensing frame is designed, which operates similarly to synchronization signal block (SSB) burst set \cite{3gpp38211}. 
    Each subframe in the environment sensing frame is beamformed towards different directions to scan the location grid, and each subframe also contains known data sequences that can be used for initial access and subsequent CKM construction.
    \item Second, we present the step-by-step ISAC signal processing algorithms based on OFDM waveform, including Periodogram, multiple signal classification (MUSIC), and Capon beamforming. 
    By using these signal processing algorithms, BS can estimate parameters such as delay, Doppler frequency, and AoA from the environment or user echoes in the monostatic ISAC scenario.
    Besides, the beam index and corresponding receive signal strength (RSS) of each path can be obtained through parameter estimation on the UE side, which can further be used to construct two types of CKM: BIM and CGM. 
    \item Finally, an ISAC-based CKM prototype system is built and its performance is tested in various quasi-static and dynamic scenarios.
    Different from \cite{exp_ckm}, the prototype does not rely on external positioning systems and achieves fully automated CKM construction for unknown environments based on ISAC. 
    The constructed CKM can provide ISAC with training-free beam alignment when LoS is obstructed.
    Compared to the location-based beam alignment, the CKM-based beam alignment provides much better performance, especially in NLoS scenarios.
    The experimental results demonstrate the feasibility and effectiveness of the ISAC-based CKM prototype systems.
\end{itemize}

The rest of this paper is organized as follows. 
Section \ref{ss_sys_mod} introduces the system model of a monostatic ISAC system based on OFDM waveform.
In Section \ref{ss_signal_process}, three mainstream ISAC signal processing algorithms are introduced, and two types of CKM are constructed.
The architecture, hardware equipment and experiment setup of the prototype are introduced in Section \ref{ss_proto_des}. 
Section \ref{ss_exp_res} shows the experimental results of the ISAC-based CKM prototype, and the comparison with the location-based benchmark is presented. 
Finally, the conclusion is drawn in Section \ref{ss_conclu}.

\section{System model}\label{ss_sys_mod}

\begin{figure}[htbp]
\centering
	{\includegraphics[width=0.43\textwidth]{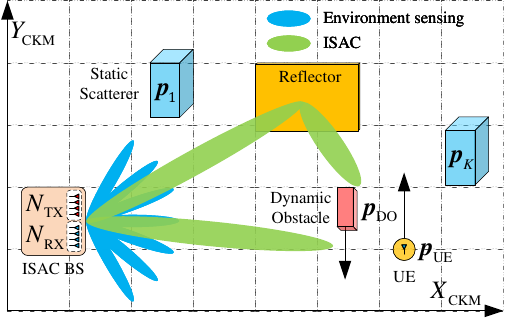}}
	\caption{Basic scenario of ISAC-based CKM.}
	\label{fig_system_sce}
\end{figure}

As shown in \mbox{Fig.~\ref{fig_system_sce}}, we consider an ISAC system with a full-duplex ISAC BS, serving a single-antenna communication UE. 
The BS is equipped with two uniform linear arrays (ULA) as transmit/receive arrays, which have $N_{\rm TX}$ and $N_{\rm RX}$ antenna elements, respectively.
The ISAC BS also senses the environment while serving the communication users.
Denote by $K=K_{\rm DO}+K_{\rm SS}$ the number of targets in the environment, including $K_{\rm DO}$ dynamic obstacles and $K_{\rm SS}$ static scatterers. 
There are two typical scenarios, namely the ISAC scenario for UEs and the environment sensing scenario for scatterers and dynamic obstacles. 
The environment sensing scenario solely involves a multi-target sensing problem, while the ISAC scenario includes both communication and multi-target sensing problems. 
In the following, we will model these problems separately.

\subsection{Communication model}
In ISAC scenario, ISAC BS transmits the downlink signal $\boldsymbol{s}(t)$ to UE, and the received signal by UE is 
\begin{equation}\label{eq_yt_UE}
\begin{aligned}
y_{\mathrm{UE}}(t)=\int_0^{\infty}{\boldsymbol{h}_{\mathrm{UE}}^{\rm T}(t,\tau )\boldsymbol{s}(t-\tau )d\tau}+w(t),
\end{aligned}
\end{equation}
where $w(t)\sim \mathcal{N}(0,\sigma^2)$ is the random Gaussian noise with variance $\sigma^2$.
$\boldsymbol{h}_{\rm UE}(t,\tau)$ denotes the channel vector between UE and ISAC BS, which can be represented as
\begin{equation}
\boldsymbol{h}_{\mathrm{UE}}(t,\tau )=\sum_{l=1}^L{\alpha _l}\boldsymbol{\beta }_{\mathrm{TX}}(\theta _l)\delta (\tau -\tau _l)e^{j2\pi \nu _lt},
\end{equation}
where $\alpha_l$, $\tau_l$, $\nu_l$, and $\theta_l$ denote the complex-valued channel coefficient, propagation delay, Doppler frequency, and the AoD of the $l$th path, respectively.
$\boldsymbol{\beta}_{\rm TX}(\theta_l)$ denotes the steering vector of the BS's ULA, where $\theta_l$ is the AoA/AoD of the $l$th path.
For an ULA with antenna spacing of $d$, its steering vector can be represented as
\begin{equation}\label{array response}
\begin{aligned}
\boldsymbol{\beta}_{\rm TX}(\theta)=\left[ 1,e^{-j\frac{2\pi d}{\lambda}\sin{ \theta}},...,e^{-j\frac{2\pi d}{\lambda}\left( N_{\rm{TX}}-1 \right)\sin{\theta} } \right] ^{\rm{T}}.
\end{aligned}
\end{equation}

\subsection{Sensing model}\label{sensing_model}
In the environment sensing scenario, ISAC BS operates as a monostatic radar, employing sensing signals to comprehensively scan the environment.
The received signal at the receive (RX) ULA of the ISAC BS can be represented as
\begin{equation}
  \bar{\boldsymbol{y}}(t) =  {{\bar{\boldsymbol{y}}_{{\rm UE}}}( t )}  + \sum\limits_{k_{\rm DO} = 1}^{{K_{{\rm{DO}}}}} {{\bar{\boldsymbol{y}}_{k_{\rm DO}}} (t )}  + \sum\limits_{k_{\rm SS} = 1}^{{K_{{\rm{SS}}}}} {{\bar{\boldsymbol{y}}_{k_{\rm SS}}}( t )} +\boldsymbol{w}(t),
\end{equation}
which consists of four components, including the reflected signal ${\bar{\boldsymbol{y}}_{{\rm UE}}}(t)$ from UE with LoS link, reflected signal ${\bar{\boldsymbol{y}}_{k_{\rm DO}}(t)}$ from $k_{\rm DO}$th dynamic obstacles, reflected signal ${\bar{\boldsymbol{y}}_{k_{\rm SS}}}(t)$ from $k_{\rm SS}$th static scatterers, and Gaussian white noise $\boldsymbol{w}(t)$.
The total number of sensing targets is $K=1+{K_{{\rm{DO}}}}+{K_{{\rm{SS}}}}$.

Since the three types of targets mentioned above have similar
sensing channels, the reflected signal from each target can be modeled in the same form:
\begin{equation}
\bar{\boldsymbol{y}}_k(t)=\int_0^{\infty}{\bar{\mathbf{H}}_{k}(t,\tau )\boldsymbol{s}(t-\tau )d\tau}+\boldsymbol{w}(t),
\end{equation}
where the channel matrix $\bar{\bf{H}}$ is a matrix of size $N_{\rm TX}\times N_{\rm RX}$:
\begin{equation}
\bar{\mathbf{H}}_k(t,\tau )=\alpha _k\boldsymbol{\beta }_{\mathrm{RX}}(\theta _k)\boldsymbol{\beta }_{\mathrm{TX}}^{\rm T}(\theta _k)\delta (\tau -\tau _k)e^{j2\pi \nu _kt}.
\end{equation}
$\alpha_{k}$ is the channel coefficient, $\tau_{k}$ and $\nu_{k}$ denote the propagation delay and the Doppler frequency of the $k$th target, respectively.
$\boldsymbol{\beta}_{\rm TX}(\theta_k)$ and $\boldsymbol{\beta}_{\rm RX}(\theta_k)$ denote the steering vector of the BS’s TX and RX antenna arrays respectively, which has similar model in (\ref{array response}).

\subsection{Signal model}
The OFDM waveform is used as the transmit signal in ISAC system due to its appealing characteristics, including the robustness to multipath fading, the exceptional sensing performance and high spectral efficiency. 
In the ISAC scenario for UEs, it is crucial to ensure uninterrupted communication while accurately sensing UE's location. 
In the environment sensing scenario, there are two objectives: tracking the locations of stationary scatterers and dynamic obstacles and utilizing beam sweeping to detect potential NLoS links between the BS and UE. 
Therefore, it is necessary for the transmitted frame to possess flexible beamforming capability to ensure continuous tracking of the UE by the BS. 
In order to solve this problem, we have designed two types of waveform based on OFDM according to different ISAC scenarios, namely environment sensing waveform (referred to as "sensing frame" for brevity) and ISAC waveform for UE (referred to as "ISAC frame"). 
Note that sensing frame is a special case of the ISAC frame.
In the following, we will focus on the sensing model of subsection \ref{sensing_model}, where both waveforms will be considered. 

The transmitted ISAC frame $\boldsymbol{s}(t)$ is represented as
\begin{equation}\label{eq_ofdmsignal}
\boldsymbol{s}(t) =\sum_{m=1}^{M_{\rm{symb}}}{\boldsymbol{s}_m(t-(m-1)T_{\rm symb})},
\end{equation}
where each frame has $M_{\rm{symb}}$ OFDM symbols and $N_{\rm{sc}}$ subcarriers, and the bandwidth of the system is $B$.
$\boldsymbol{s}_m(t)$ is the $m$th transmitted OFDM symbol:
\begin{equation}\label{eq_ofdmsignal_symb}
\begin{aligned}
\boldsymbol{s}_m(t) =\boldsymbol{f}_{\rm{TX},m}\sum_{n=1}^{N_{\mathrm{sc}}} b_{n,m}e^{j2\pi \left( n-1 \right) \Delta f\left( t-T_{\rm{CP}} \right)}{\rm rect}\left(\frac{t}{T_{\rm symb}}\right),
\end{aligned}
\end{equation}
where $t \in \left[0,T_{\rm symb} \right)$, $b_{n,m}$ denotes the information-bearing symbol on the $n$th subcarrier of the $m$th  OFDM symbol, $T_{\rm{CP}}$ is the cyclic prefix (CP) duration, $\Delta f$ is the subcarrier spacing and $T_{\rm{symb}}$ denotes the time duration of each OFDM symbol including CP.
The system is designed to employ beamforming vector $\boldsymbol{f}_{\rm{TX},m}$ for the single-RF chain phased array during transmission of the $m$th symbol. 

For the sensing frame $\tilde{\boldsymbol{s}}(t)$, the symbols $\{b_{m,n}\}$ in (\ref{eq_ofdmsignal_symb}) are replaced with a known sequence $\{\tilde{b}_{m,n}\}$.
In this paper, the Zadoff-Chu sequence\cite{3gpp36211} is selected for its superior radar performance resulting from significantly lower peak to average power ratio (PAPR) after OFDM modulation \cite{zrq2019zc}. 
The sensing frame is divided into $Q=\lfloor M_{\rm{symb}}/M_{\rm{q}}  \rfloor $ subframes, with each subframe containing $M_{\rm{q}}$ OFDM symbols.
The beamforming vector of $m$th symbol $\boldsymbol{f}_{{\rm TX},m}$ is switched in each subframe, thus the total number of beams is $Q$.
\begin{equation}\label{eq_bf_vec}
\boldsymbol{f}_{{\rm TX},m}=\left[ 1,e^{j\frac{2\pi}{Q}\lfloor \frac{m}{M_{\rm{q}}} \rfloor},...,e^{j\frac{2\pi}{Q}\lfloor \frac{m}{M_{\rm{q}}} \rfloor \left( N_{{\rm TX}}-1 \right)} \right] ^{\mathrm{T}},
\end{equation}
where $\lfloor{.}\rfloor$ represents the rounding down of number.
For the $q$th beam, its corresponding beamforming angle is $\theta_{q}$, which divides the azimuth angular space into $Q$ parts:
\begin{equation}
\begin{aligned}
        {\theta _q} = \left\{ {
\begin{matrix}
           {\arcsin (\frac{2q}{Q})} & {\frac{2q}{Q} \in \left[ {0,1} \right]}  \\
          {\arcsin (\frac{2q}{Q} - 2)} & {\frac{2q}{Q} \in \left( {1,2} \right]}  \\ 
 \end{matrix}
 } \right.
\end{aligned}.
\end{equation}

\begin{figure}[htbp]
\centering
	{\includegraphics[width=0.48\textwidth]{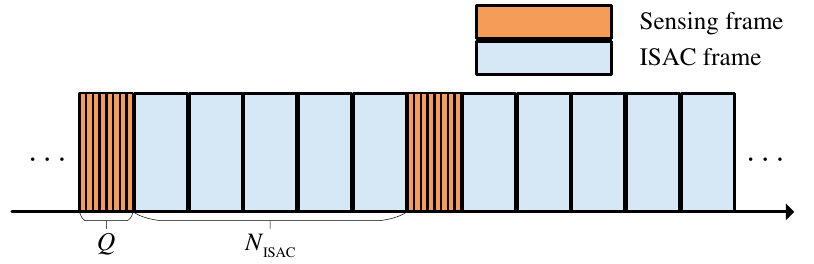}}
	\caption{Frame structure for different ISAC scenarios.}
	\label{fig_isac_sequence}
\end{figure}

The employed OFDM frame structure varies between the two scenarios mentioned above, as shown in \mbox{Fig.~\ref{fig_isac_sequence}}, where $Q$ denotes the number of beams used for environment sensing. 
$N_{\rm ISAC}$ denotes the interval of environment sensing frame, indicating that the BS performs a beam scanning every $N_{\rm ISAC}$ ISAC frames.

The next step involves extracting channel matrix from the echo of ISAC frame or sensing frame, in order to further sense UE or scatterers. 
The received signal $\boldsymbol{y}(t)$ of BS is then divided into $M_{\rm{symb}}$ blocks with equal block duration $T_{\rm{symb}}$, with the $m$th block given by
\begin{equation}\label{segmentation}
\begin{aligned}
\bar{\boldsymbol{y}}_m(t) =\bar{\boldsymbol{y}}( t+(m-1)T_{\rm symb}) \mathrm{rect}\left( \frac{t}{T_{\mathrm{symb}}} \right).
\end{aligned}
\end{equation}
Then, the receive beamforming and CP removal are then performed on block $\bar{\boldsymbol{y}}_m(t)$, yielding
\begin{equation}\label{recv_bf}
\begin{aligned}
y_m(t)=\boldsymbol{f}_{\mathrm{RX},m}^{\mathrm{T}}\bar{\boldsymbol{y}}_m\left( t+T_{\mathrm{CP}} \right) 
\mathrm{rect}\left( \frac{t}{T_{\mathrm{O}}} \right),
\end{aligned}
\end{equation}
where $T_{\mathrm{O}}=T_{\mathrm{symb}}-T_{\rm{CP}}$ is the OFDM symbol duration without CP.
    Similar to (\ref{eq_bf_vec}), the receive beamforming vector is denoted by $\boldsymbol{f}_{\rm{RX},m}$.
Assuming that CP duration is larger than the maximum delay, i.e., $T_{\mathrm{CP}}>\max \left( \left\{ \tau _k \right\} _{k=1}^{K} \right) $, (\ref{recv_bf}) can be further written as
\begin{equation}\label{recv_bf_2}
\begin{aligned}
&y_m(t) =\sum_{k=1}^K{\left\{ \gamma _{k,m}\sum_{n=1}^{N_{\mathrm{sc}}}{b_{n,m}e^{j2\pi \left( n-1 \right) \Delta f\left( t-\tau _k \right)}} \right.}
\\
&\left. e^{j2\pi \nu _k\left( t+\left( m-1 \right) T_{\mathrm{symb}} \right)}\mathrm{rect}\left( \frac{t}{T_O} \right) \right\} +w_m(t),
\end{aligned}
\end{equation}
where $\gamma _{k,m}=\alpha _k\boldsymbol{f}_{\mathrm{RX},m}^{\mathrm{T}}\boldsymbol{\beta }_{\mathrm{RX}}(\theta _k)\boldsymbol{\beta }_{\mathrm{TX}}^{T}(\theta _k)\boldsymbol{f}_{\mathrm{TX},m}$ and $w_m(t)$ is the additive Gaussian noise. 
Signals $\{y_m(t)\}_{m=1}^{M_{\rm{symb}}}$ are then sampled at an interval of $1/B$, and then $N_{\rm sc}$-point fast Fourier transform (FFT) is performed on each symbol to obtain a discrete time-frequency domain matrix:
\begin{equation}\label{eq_block_fft}
\begin{aligned}
    {{\bf Y}_{n,m}} = \sum\limits_{k = 1}^K {\gamma _{k,m}}b_{n,m}{e^{j{\omega _{k,n,m}}}}+{\bf W}_{n,m},
\end{aligned}
\end{equation}
where ${\omega _{k,n,m}}$ denotes the phase shift caused by delay and Doppler effect of $m$th symbol and $n$th subcarrier from $k$th target:
\begin{equation}\label{eq_dealy_dop}
{\omega _{k,n,m}} = 2\pi \left( {{\nu_k}\left(m-1\right){T_{\rm{symb}}} - {\tau _k}\left(n-1\right)\Delta f} \right).
\end{equation}

Since the BS knows all the transmitted data $\textbf{b}$, it can be directly removed in $\bf Y$ by element-wise division to obtain a channel matrix of size $M_{\rm{symb}} \times N_{\rm{sc}}$:
\begin{equation}\label{eq_y/b}
\begin{aligned}
    {{\bf H}_{n,m}} = \frac{{\bf Y}_{n,m}}{{b_{n,m}}} =\sum\limits_{k = 1}^K {\gamma _{k,m}}{e^{j{\omega _{k,n,m}}}}+{\bf W}'_{n,m}.
\end{aligned}
\end{equation}
The matrix ${{\bf H}}$ and ${{\bf Y}}$ can be further sliced for ${Q}$ beams, such as $\{\mathbf{H}_{q}\}_{q=1}^{Q}$, where the $(n,m)$th element of the channel matrix for the $q$th beam is
\begin{equation}\label{Beam_Matrix}
 \left[ \mathbf{H}_{q} \right] _{n,m}=\mathbf{H}_{n,(q-1)M_{\rm{q}}+m}.
\end{equation}
${{\bf Y}}$ can be represented as $\{\mathbf{Y}_{q}\}_{q=1}^{Q}$, similarly.

\section{Signal processing}\label{ss_signal_process}
In order to extract sensing information from the received signals, we apply radar signal processing algorithms to process the channel matrices $\{\mathbf{H}_{q}\}_{q=1}^{Q}$ in (\ref{Beam_Matrix}). 
The ISAC-based CKM prototype system considers Periodogram \cite{2014ofdmradar}, MUSIC \cite{music} and Capon beamforming \cite{capon} as the signal processing algorithms, each of which has its own advantages and disadvantages. 
Therefore, the algorithm can be flexibly chosen according to the application scenario. 

\subsection{Periodogram}
For ISAC range sensing, IDFT with $N_{\rm{IDFT}}$ points is applied to each column of $\mathbf{H}_{q}$ in (\ref{Beam_Matrix}), and the $q$th element of the periodogram is calculated as
\begin{equation}\label{eq_per_t}
	\begin{aligned}
	{P_{q}^{\rm Per}} = \sum\limits_{m = 1}^{M_{\rm{q}}} {\frac{1}{N_{\rm{sc}}M_{\rm{q}}}{{\left| {{\rm{IDFT}}\left(\left[ \mathbf{H}_{q} \right] _{:,m} \right)} \right|}^2}},
 	\end{aligned}
\end{equation}
where $\left[ \mathbf{H}_{q} \right] _{:,m}$ denotes the $m$th column of matrix $\left[ \mathbf{H}_{q} \right]$. 
The complete range-angle spectrum is denoted by: 
\begin{equation} 
    {\bf P}^{\rm Per}
= \left[ {P_{1}^{\rm Per}},...,{P_{Q}^{\rm Per}} \right]
 \in \mathbb{R} ^{N_{{\rm{IDFT}}} \times \mathrm{Q}}.
\end{equation}

The peak of $ {P}^{\rm Per}_{q}$ is $\hat{\boldsymbol{n}}_{\rm IDFT}$, which indicates the delay of the corresponding targets at angle ${\boldsymbol{\theta}_q}$ and then the ranges can be estimated by
\begin{equation}
\begin{aligned}
    \hat{\boldsymbol{r}}_q &= \frac{\hat{\boldsymbol{n}}_{\rm IDFT} \cdot c}{2\Delta f{N_{{\rm{IDFT}}}}},
\end{aligned}
\end{equation}
where $c$ denotes the speed of light.

Since the periodogram can be obtained based on FFT, it has the lowest computational complexity and highest refresh rate. 
However, its range resolution is limited by the signal bandwidth, i.e.,
\begin{equation}
\Delta r=\frac{c}{2B}.
\end{equation}

\subsection{MUSIC}
In order to achieve super resolution, the MUSIC algorithm can be used for parameter estimation. 
However, the coherence of echoes from multiple targets will degrade the algorithm's performance.
Therefore, we have to apply the modified spatial smoothing preprocessing (MSSP) \cite{Superresolution_techniques_for_time_domain} and forward–backward correlation matrix (FBCM) \cite{Statisical_and_Adaptive_Signal_Processing} in subcarrier dimension to eliminate the ambiguity of range peaks.

For the channel matrix $\mathbf{H}_{q}$ of size $N_{\rm sc} \times M_{\rm{symb}}$, ${N_{{\rm{sub}}}}$ submatrices of size $L \times {N_{{\rm{symb}}}}$ are selected in all the ${N_{{\rm{sc}}}}$ subcarriers, where ${N_{{\rm{sub}}}}{= }{N_{{\rm{sc}}}} - L + 1$, $L = \rho {N_{{\rm{sc}}}}$ and $\rho=0.4 $ is the smoothing constant.
The channel matrix after spatial smoothing can be expressed as:
\begin{equation} 
    {{\bf H}_{q}'} 
= \left[ {
\begin{matrix}
   {\left[ \mathbf{H}_{q} \right] _{1,:}} & ... & {\left[ \mathbf{H}_{q} \right] _{n_{\rm sub},:}}  \\ 
   \vdots  &  \ddots  & \vdots \\
   {\left[ \mathbf{H}_{q} \right] _{L,:}} & ... & {\left[ \mathbf{H}_{q} 
\right] _{n_{\rm sub}+L,:}}  \\ 
 \end{matrix} } \right]^{\rm T}
\end{equation}
where $\left[ \mathbf{H}_{q} \right] _{n,:}$ denotes the $n$th row of $\mathbf{H}_{q}$, and $n_{\rm sub} \in \left[ {1,{N_{{\rm{sub}}}}} \right]$. 

The covariance matrix of ${{\bf H}_{q}'} $ can be obtained by using FBCM as:
\begin{equation}
{\bf R}_{q} = \frac{1}{2}\left( {{\bf R}_{q}'}  + {\bf{U}}{{\left( {{\bf R}_{q}'} \right)}^*}{\bf{U}} \right),
\end{equation}
where ${\bf{U}}$ is the $L \times L$ all-zero matrix except the $l$-th antidiagonal entry being 1, and 
\begin{equation}
     {{\bf R}_{q}'}  = \frac{{{\bf H}_{q}'} {{{\bf H}_{q}'}^{\rm{H}}}}{N_{{\rm{sub}}}}.
 \end{equation}

Then, the eigenvalue decomposition is performed on ${\bf R}_{q}$:
\begin{equation}\label{eq_rtau}
	\begin{aligned}
{\bf R}_{q} = {\bf{E}}_{{\rm s},{q}} {\bf{\Lambda}}_{{\rm s},{q}} {{\bf{E}}_{{\rm s},{q}}^{\rm{H}}} + {\bf{E}}_{{\rm n},{q}} {\bf{\Lambda }}_{{\rm n},{q}} {{\bf{E}}_{{\rm n},{q}} ^{\rm{H}}},
	\end{aligned}
\end{equation}
where ${\bf{\Lambda}}_{{\rm s},{q}}$ denotes the diagonal matrix with respect to the $N_{\rm s}$ largest eigenvalues, and ${\bf{E}}_{{\rm s},{q}}$ and ${{\bf{E}}_{{\rm n},{q}}}$ denote the signal subspace and noise subspace, respectively. 
Therefore, the MUSIC spectrum can be obtained as:
\begin{equation}
	\begin{aligned}
P_{q}^{{\rm{MUSIC}}}\left( \tau  \right) = \frac{1}{{{\boldsymbol{x}(\tau)^{\rm{H}}}{\bf{E}}_{{\rm n},{q}} { {\bf{E}}_{{\rm n},{q}}^{\rm{H}}}{\boldsymbol{x}(\tau)}}}
\label{eqmusic}
	\end{aligned}
\end{equation}
where the $n$th element of the steering vector $\boldsymbol{x}(\tau)$ can be expressed as $x_{n}(\tau)  = {e^{ - j2\pi n\tau\Delta f }}$.
Denote the searching interval on the delay dimension as $\Delta\tau$, and then $\tau$ can be expressed as $\tau =u\Delta\tau$, where $u \in \{0,\ldots,U-1\}$ denotes the index of the delay grid. 
The range-angle spectrum can then be expressed as:
\begin{equation}\label{music}
    {\bf P}^{{\rm{MUSIC}}} = [{P}_{1}^{\rm{MUSIC}},...,{P}_{Q}^{\rm{MUSIC}}]\in \mathbb{R} ^{U\times \mathrm{Q}}, 
\end{equation}
where ${P}_{q}^{\rm{MUSIC}}=[P_{q}^{{\rm{MUSIC}}}(0),\ldots,P_{q}^{{\rm{MUSIC}}}((U-1)\Delta\tau)]^\mathrm{T}$.
   
MUSIC has the best resolution performance, but the spectrum search in equation (\ref{eqmusic}) leads to extremely high complexity. 
Additionally, the principle of the MUSIC algorithm also causes its estimation results to be affected by other targets in the environment, meaning that the MUSIC spectrum cannot reflect the true reflected signal strength of the target.

\subsection{Capon beamforming}
MUSIC can offer a high-resolution estimation of channel parameters. 
However, the peaks of its spectrum fail to accurately reflect the true power of the targets. 
Hence, the Capon beamforming\cite{capon} is considered as a reliable compromise solution. 

The spectrum of Capon beamforming can be obtained as:
\begin{equation}\label{eq_capon}
	\begin{aligned}
{P}_{q}^{{\rm{Capon}}}(\tau) = \frac{1}{{\boldsymbol{x}(\tau)^{\rm{H}}}{\bf R}_{q}^{-1}{\boldsymbol{x}(\tau)}}.
	\end{aligned}
\end{equation}
Similar to (\ref{music}), the range-angle spectrum of Capon beamforming can also be represented as a matrix 
${\bf P}^{\rm{Capon}} \in \mathbb{R} ^{U\times \mathrm{Q}}$.

\subsection{Clutter rejection and target tracking}
So far, the signal processing algorithms that we have discussed are based on ideal single-target scenarios for estimating target positions. 
However, in real environments, due to the interference of clutter, the number of echoes usually does not equal to the number of targets, making it challenging to determine whether measurement data is from a tracked target or a unwanted clutter. 
In order to distinguish targets from complex environments, clutter rejection and clustering are necessary. 
For the three signal processing algorithms mentioned earlier, Periodogram and Capon beamforming reflect the true power of targets.
In addition, the algorithm's complexity is also an important factor. 
Monte Carlo simulation is conducted for the three algorithms using the parameters in Table \ref{NR Parameters} and their average execution time and root mean squared error (RMSE) of range estimation is summarized, as shown in Table \ref{alg_simu}.
The result shows that the average execution time of Capon beamforming is almost the same as MUSIC, and the range estimation performance of Capon beamforming is better than that of the other two algorithms.
Besides, Capon beamforming has the characteristic of reflecting the true power of targets.
Therefore, we finally choose Capon beamforming as the signal processing algorithm in the subsequent experiments, balancing the algorithm performance and complexity.

\begin{table}[hbtp]
\begin{center}
\caption{Performance of three algorithms in Monte Carlo simulation.}
			\begin{tabular}{|c|c|c|c|}
				\hline
                Algorithm       & Average execution time  &   RMSE          \\ \hline
			    Periodogram     & \SI{0.0062}{\second}    &   \SI{3.5329}{\meter}           \\ \hline
				MUSIC           & \SI{0.5869}{\second}    &   \SI{2.7912}{\meter}              \\ \hline
                Capon beamforming  & \SI{0.5642}{\second} &   \SI{2.4464}{\meter}               \\ \hline
			\end{tabular}
		\label{alg_simu}
	\end{center}
\end{table}

Thus clutter map rejection can be used by subtracting the range-angle spectrum of static environment $\dot{\bf P}$ from the sensing result $\bf P$, thereby distinguishing dynamic objects.
The sensing result that have undergone clutter rejection is represented as
\begin{equation}
    {\tilde{{\bf P}}} = {\bf P} - \dot{\bf P}.
\end{equation}

There have been many excellent works on clustering algorithms, such as k-means \cite{k_means}, mean shift \cite{mean_shift}, and density based spatial clustering of applications with noise (DBSCAN) \cite{DBSCAN}. 
After clutter rejection of range-angle spectrum, the range $\tilde{\boldsymbol{r}}$ and angle $\tilde{\boldsymbol{\theta}}$ of the target cluster center is extracted as
\begin{equation}
    {\boldsymbol{p}} = \left({\tilde{\boldsymbol{r}}}\cos{\tilde{\boldsymbol{\theta}}},{\tilde{\boldsymbol{r}}\sin{\tilde{\boldsymbol{\theta}}}} \right).
\end{equation}

Additionally, to further enhance the reliability of multi-target tracking, the adoption of joint probabilistic data association (JPDA) algorithm \cite{JPDA} is recommended. 
The JPDA is a target tracking algorithm based on Bayes filtering, which estimates the state of targets by jointly considering all possible data associations.

\begin{figure*}[!ht]
\centering
	{\includegraphics[width=0.80\textwidth]{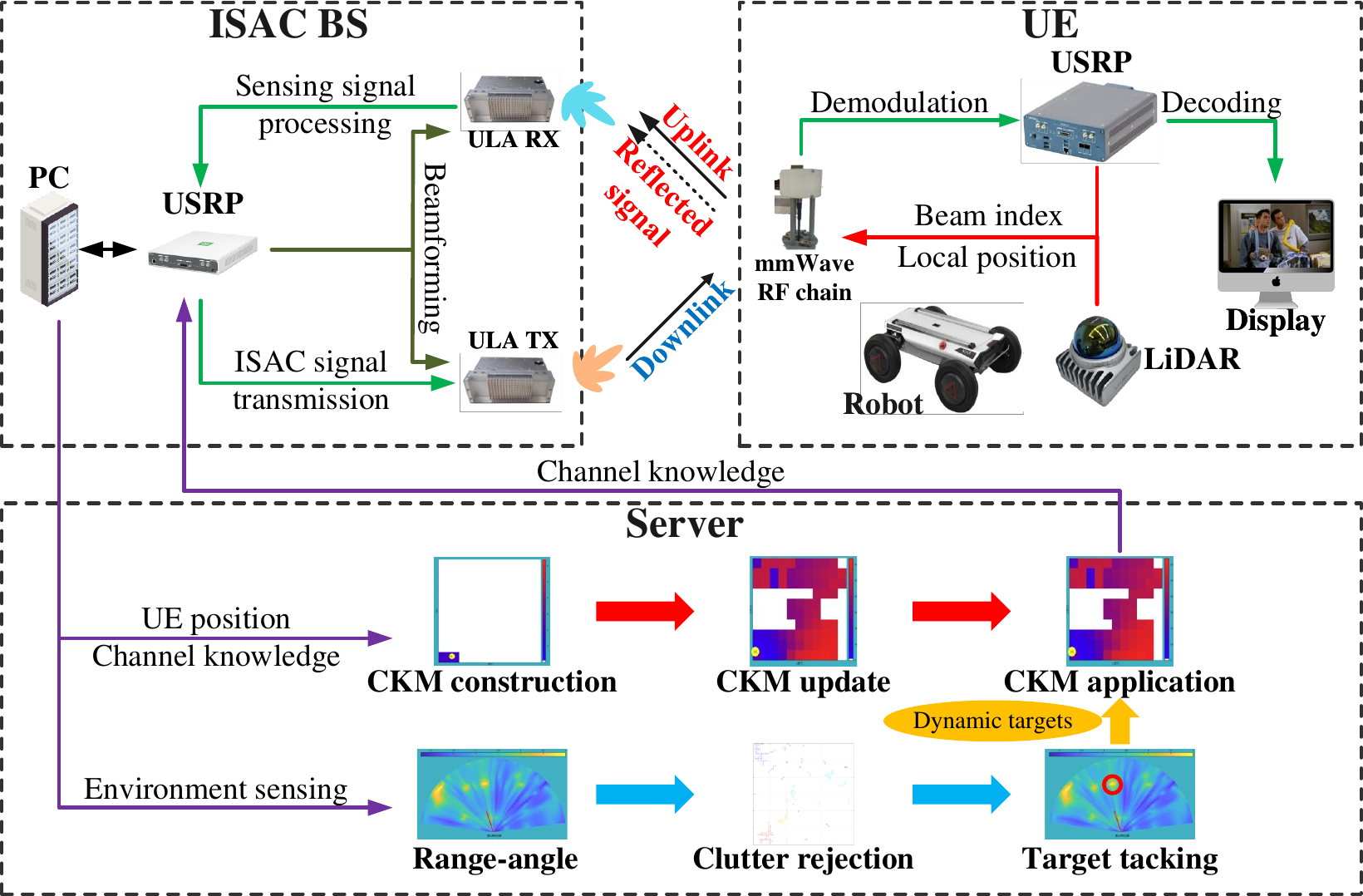}}
	\caption{System structure of ISAC-based CKM prototype.}
	\label{fig_structure}
\end{figure*}

\subsection{CKM construction}
CKM directly stores the indices of multiple transmit beams at each interested location, as well as their corresponding RSS. 
Therefore, the constructed types of CKM includes BIM and CGM.  
We divide the region of interest $\bf \Psi$ into $I=X_{\text{CKM}} \times Y_{\text{CKM}}$ grids, denoted as ${\bf \Psi} = \left\{\psi_1,...,\psi_I \right\}$. 
To construct the CKM, UE moves within the region of interest and estimates the parameters of the downlink sensing frame $\tilde{\boldsymbol{s}}(t)$ from ISAC BS until the entire area is traversed.
During this process, location $\boldsymbol{p}$ will be allocated to the corresponding grid by $\boldsymbol{p}_{{\psi}_i}={\rm round}(\boldsymbol{p})$.
Where the function $\rm round(\cdot)$ represents the rounding function, which rounds the location $\boldsymbol{p}$ to the center $\boldsymbol{p}_{{\psi}_i}$ of its nearest grid.
Then, the CKM can be expressed as:
\begin{equation}\label{ckm}
    \left\{ {{\psi}_i;\left( {{F}}_i ,{{A}}_i\right) }\right\}_{i=1}^I,
\end{equation}
where ${F}_i$ and ${{A}}_i$ denote the set of candidate beam indices and their corresponding RSS in the $i$th grid, respectively.
Similar to equation (\ref{eq_block_fft}), the UE can obtain the time-frequency domain matrix $\tilde{\bf Y}$ of the downlink signal $\tilde{\boldsymbol{y}}$. 
For each sensing beam $\tilde{s}_{q}$, its received RSS by the UE is represented as ${A}_i = \left[a_1,...,a_Q \right]$, where
\begin{equation}
    a_{q} = 10 \lg \left( \frac{\Vert{\bf Y}_{\rm q}\Vert^2_{\rm F}}{M_{{\rm{q}}}}\right).
\end{equation}
Then, sort the RSS in descending order and record $N_{\text {max}}$ strongest beam indices. 
The set of strongest beam indices is represented as
\begin{equation}
    F_i = \left\{ \boldsymbol{f}_{i,1},\boldsymbol{f}_{i,2},...,\boldsymbol{f}_{i,N_{\rm max}} \right\}.
\end{equation}

\section{Prototyping design}\label{ss_proto_des}
\subsection{Prototype architecture}\label{arch}
The architecture of ISAC-based CKM prototype is shown in \mbox{Fig.~\ref{fig_structure}}, which consists of three parts: ISAC BS, UE and cloud server. 

The ISAC BS contains a Universal Software Radio Peripheral (USRP), two mmWave phased arrays and a personal computer (PC).
The USRP serves as the fundamental element of ISAC BS, functioning similarly to a baseband unit (BBU) of the BS and assuming responsibility for the majority of baseband signal processing tasks, including constellation mapping and OFDM modulation. 
The mmWave phased array adopts a $1 \times 16$ ULA analog beamforming architecture which has a single channel of intermediate frequency and mmWave up-and-down converter. 
Beamforming can be achieved by externally controlling the phase shifters, which enables symbol-level flexible beamforming via direct field-programmable gate array (FPGA) control over the mmWave phased array. 

The PC serves as the host computer of ISAC BS, performs essential functions such as radar signal processing, data interaction, and visualization. 
On one hand, it acquires undemodulated physical layer data from FPGA to execute parameter estimation algorithms. 
On the other hand, it uploads user's feedback channel knowledge data along with its positioning information to cloud server and retrieves location-based channel knowledge to facilitate pilot-free mmWave communication. 

The UE contains a mmWave RF link, USRP, light detection and ranging (LiDAR) device, and a mobile robot. 
The mmWave RF link consists of an omnidirectional mmWave antenna and a mmWave up-and-down converter to convert the received signal to intermediate frequency and transmit it to the USRP. 
Similarly, the USRP performs tasks such as OFDM demodulation and channel parameter estimation. 
The UE system is mounted on a mobile robot along with the LiDAR and continuously moves in the environment. 

The server is also a PC used to build CKM or feedback relevant information from CKM to BS based on location information.

\subsection{Hardware equipment}
\begin{figure*}[!hb]
	\centering
        \subfigure[Hardware structure of ISAC BS.]{\label{fig_exp_bs}
        \includegraphics[width=0.36\textwidth]{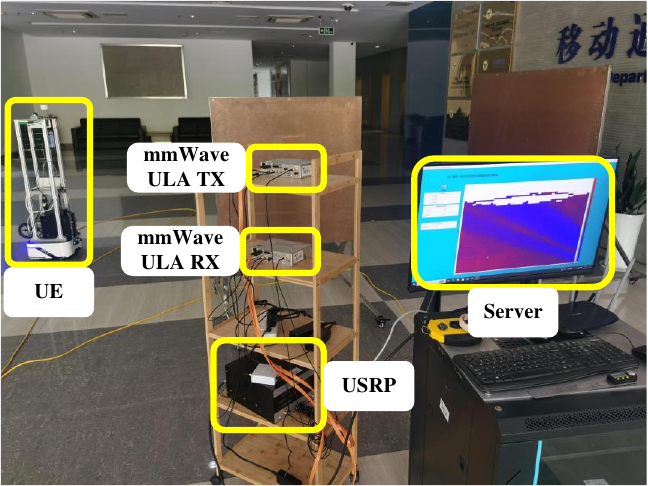}
		}
	\hfil
	\subfigure[Hardware structure of UE.]{\label{fig_exp_ue}
        \includegraphics[width=0.2028\textwidth]{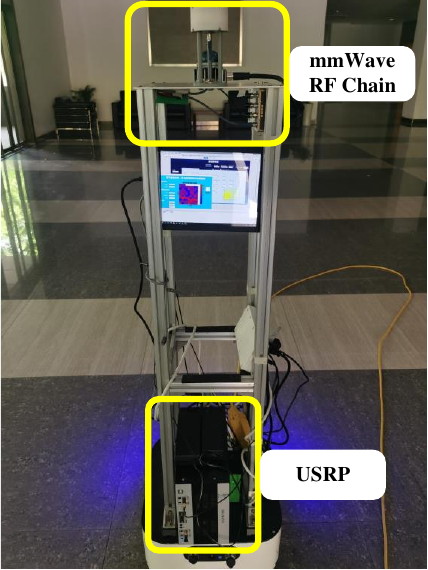}
        } 
        \hfil
        \subfigure[Measurement equipment on UE.]{\label{fig_exp_equ}
        \includegraphics[width=0.36\textwidth]{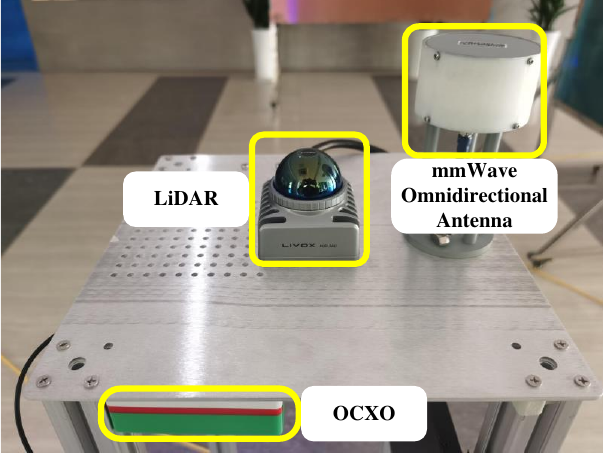}}
        
    \caption{Hardware structure of ISAC-based CKM prototype.}
\end{figure*}
As mentioned in subsection \ref{arch}, ISAC BS and UE use USRP as the main component, and mmWave phased array and mmWave RF chain are used as RF transceiver components respectively. 
The mobile robot of UE is also equipped with a LiDAR, which utilizes a single board computer (SBC) to assist simultaneous localization and mapping (SLAM). 
The specific hardware equipment selection and capabilities are as follows:

\subsubsection{USRP} 
The NI USRP-2974 is a Software Defined Radio (SDR) stand-alone device built with a FPGA and an onboard processor for rapidly prototyping high-performance wireless communication systems. 
With a real time bandwidth of up to $\SI{160}{\mega\hertz}$, two daughter boards covering 10MHz to 6GHz are utilized to perform analog up-and-down conversion. 
The baseband signal sampling/interpolation and digital up-and-down conversion are realized rapidly on the Xilinx Kintex-7 410T FPGA, while the Intel i7 onboard processor is responsible for waveform related operations like modulation/demodulation. 
Configured by the NI LabView Communication program, USRP-2974 can be operated as a transmitter to generate required signals flexibly and a receiver to process signals efficiently. 

\subsubsection{mmWave phased array}
The mmPSA TR16-1909 is used as a phased antenna array for mmWave communication systems, which is a 16-element phased array and provides vertical polarization service. 
mmPSA TR16-1909 is cable-connected to USRP, to which the control frame is transmitted from the NI LabVIEW Communications program to realize exhaustive beam sweeping. 
The device supports time division duplex (TDD) communication, whose operating frequency band is $\SI{27}{\giga\hertz}$ to $\SI{29}{\giga\hertz}$. 
The phase array provides the fast and normal modes in phase control. 
In the fast mode, 16 phase modules can be set independently, and the setting accuracy of each phase module is 5 bits.

\subsubsection{mmWave RF chain}
mmWave RF chain is composed of a mmWave omnidirectional antenna and a mmWave frequency converter, with the model number of the mmWave frequency converter being mmFE-28-4CM. 
mmFE-28-4CM is a $\SI{28}{\giga\hertz}$ TDD 4-channel mmWave transceiver that integrates mmWave low-noise amplifiers and mmWave power amplifiers. 
It has excellent noise characteristics and transmission signal accuracy, making it suitable for applications such as NR mmWave communication testing, instrument measurement, and production line testing. 
One channel of the transceiver is connected to the mmWave omnidirectional antenna, which operates in the frequency range of $\SI{18}{\giga\hertz}$ to $\SI{30}{\giga\hertz}$. 
The antenna has an omnidirectional radiation pattern in the horizontal direction to prevent a decrease in received power caused by UE rotation. 

\subsubsection{LiDAR and SBC}
The measurement equipment is shown as \mbox{Fig.~\ref{fig_exp_equ}}.
The model of the LiDAR is MID-360, which has the characteristics of small size and flexible deployment. 
It can scan 200,000 points per second with a field of view (FOV) of $\SI{360}{\degree} \times \SI{59}{\degree}$. 
The minimum detection range is as low as $\SI{10}{\centi\meter}$, and under conditions of 80\% reflectivity, the maximum detection range can reach $\SI{70}{\meter}$. 

In order to achieve SLAM for obtaining UE's location, we have also equipped the LiDAR with an additional SBC to run robot operating system (ROS) on it. 
The chosen system on chip (SOC) model is RK3588, manufactured by $\SI{8}{\nano\meter}$ process technology, featuring quad-core Cortex-A76 and quad-core Cortex-A55, possessing a computility of 6TOPS.

\subsection{Experiment setup}
\begin{figure}[htbp]
\centering
	{\includegraphics[width=0.43\textwidth]{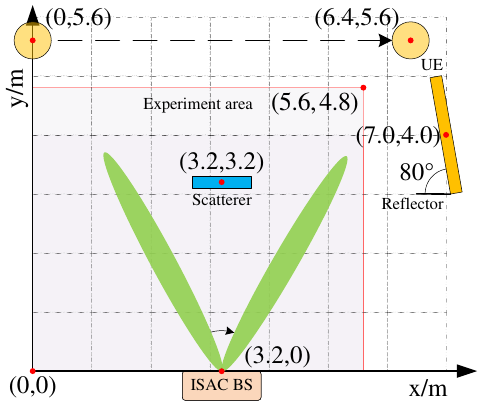}}
	\caption{Experiment scenario.}
	\label{exp_scenario}
\end{figure}

The experiment scenario is shown as \mbox{Fig.~\ref{exp_scenario}}, where the ISAC BS is placed at a distance of 3.2m from the origin on the x-axis. 
There is a metal scatterer with a width of $\SI{1}{\meter}$ located exactly 3.2m in front of the ISAC BS. 
Additionally, there is a metal reflector with a width of $\SI{2}{\meter}$ positioned and an angle of $\SI{80}{\degree}$ at point $(7.0,4.0)$ to provide a stable NLoS path for the UE.
We designed two functional validation experiments to verify the inherently mutualistic relationship between ISAC and CKM: ISAC for CKM construction and CKM application for ISAC.

The corresponding waveform parameters for subsequent prototyping experiments are specified in Table \ref{NR Parameters}. 
\begin{table}[hbtp]
\begin{center}
\caption{Parameters for ISAC-based CKM prototype.}
			\begin{tabular}{|c|c|}
				\hline
                carrier frequency, ${f_c}$                & \SI{27.5}{\giga\hertz}      \\ \hline
				bandwidth, $B$                            & \SI{80}{\mega\hertz}     \\ \hline
				subcarrier spacing, $\Delta f$           & \SI{78.125}{\kilo\hertz}\\ \hline
				number of subcarriers, $N_{{\rm{sc}}}$              & 1024                      \\      \hline
                frame length, $T_{{\rm{frame}}}$      & \SI{16.432}{\milli\second}                      \\ \hline
                symbol length, $T_{{\rm{symb}}}$      & \SI{16}{\micro\second}                      \\ \hline
                number of symbols per frame, $N_{{\rm{symb}}}$      & 1027                      \\ \hline 
                antenna spacing, $d$ & \SI{5.4}{\milli\meter} \\ \hline 
                element number of ULA, $N_{\rm ULA}$& 16 \\ \hline
                constellation                                  & QPSK
                  \\ \hline
                   
			\end{tabular}
		\label{NR Parameters}
	\end{center}
\end{table}

\subsubsection{ISAC for CKM construction}
\begin{figure}[htbp]
\centering
	{\includegraphics[width=0.43\textwidth]{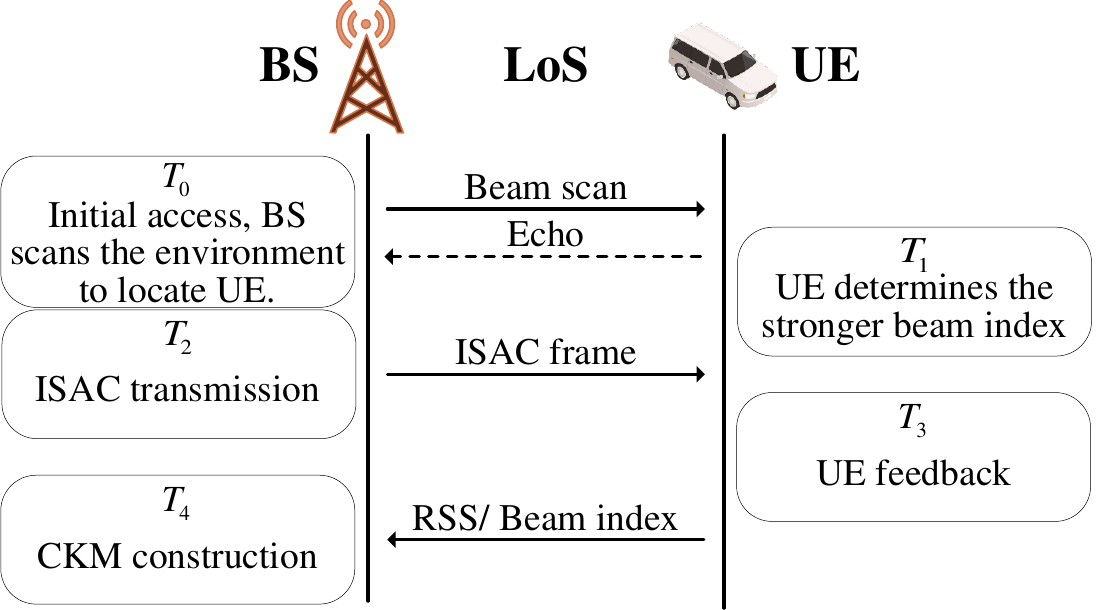}}
	\caption{CKM construction program.}
	\label{fig_prog_cons}
\end{figure}

To verify the feasibility of ISAC-based CKM construction, in the first experiment, we construct the CKM for the experiment area in \mbox{Fig.~\ref{exp_scenario}}. 
Based on the principles described in section \ref{ss_signal_process}, ISAC BS can determine the absolute location of UE $p_{\rm UE}$ under LoS condition. 
The interactive program for CKM construction is shown in \mbox{Fig.~\ref{fig_prog_cons}}:
\begin{itemize}
    \item[$T_0$:] ISAC BS periodically conducts environment sensing with the sensing frames and performs initial access to the UE within the experiment area to obtain its location. 
    \item[$T_1$:] UE leverages the sensing frames for beam training and obtains the strongest beam indices. 
    \item[$T_2$:] After establishing an ISAC link under the LoS cases, the ISAC BS uses ISAC frame to transmit data to UE while sensing the location of it. 
    \item[$T_3$:] UE reports the strongest beam indices and their corresponding RSS to the ISAC BS. 
    Besides, when UE enters the NLoS area, the location of the UE is also reported.
    \item[$T_4$:] ISAC BS stores the received feedback information with the current location of the UE on the server, and the CKM construction of one grid is completed.
\end{itemize}

The UE moves from the origin of \mbox{Fig.~\ref{exp_scenario}}, traversing all the grids from left to right and bottom to top. 
Its movement continues until it reaches the upper right corner of the area, actively avoiding scatterers in each corresponding grid.
This process signifies the successful completion of CKM construction as the UE covers all grids within the experiment area.
Notably, to accommodate larger movement areas for subsequent experiments, the actual CKM construction area is larger than that shown in \mbox{Fig.~\ref{exp_scenario}}. 
For better visual presentation, blank gaps at the boundaries of the practical environment have been trimmed in displaying subsequent CKM.

\subsubsection{CKM application for ISAC}
To verify the benefit brought by applying CKM for ISAC, in the second experiment, CKM-based training-free beam alignment is applied for mobile UE.
We will compare the performance of location-based beam alignment and CKM-based beam alignment. 
In \mbox{Fig.~\ref{exp_scenario}}, the UE is moved at a constant speed from point $(0,5.6)$ to point $(6.4,5.6)$, and the variation of UE-side RSS is compared between ISAC BS using location-based beam alignment and CKM-based beam alignment. 
Here, location-based beam alignment refers to ISAC BS steering the beamforming vector to the corresponding angle based on the feedback of UE's location.

\begin{figure}[htbp]
\centering
	{\includegraphics[width=0.43\textwidth]{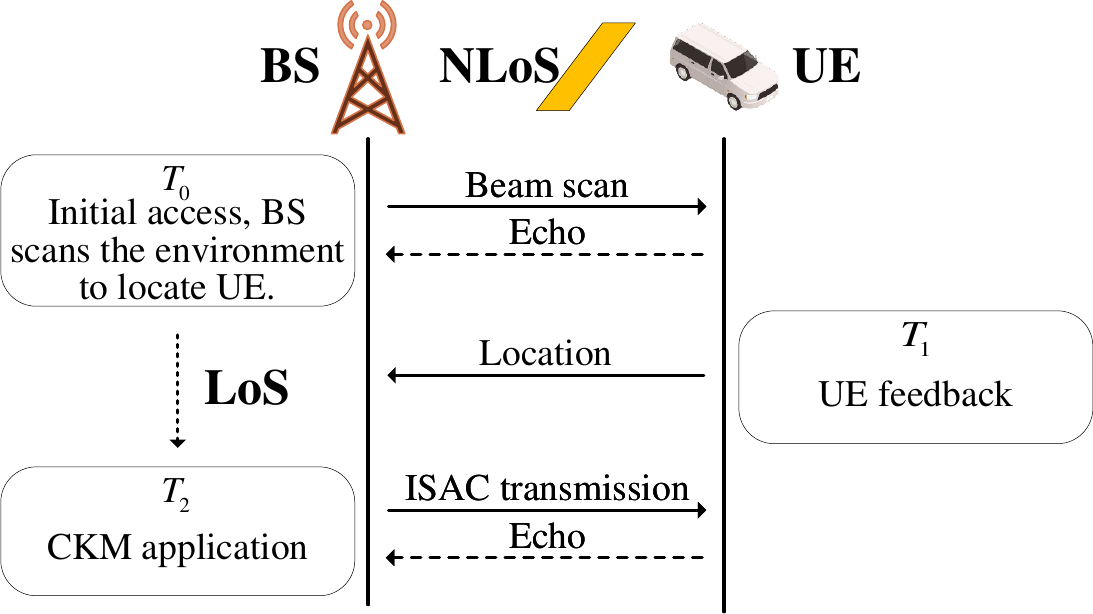}}
	\caption{CKM application under LoS/NLoS condition.}
	\label{program_cons_nlos}
\end{figure}

The interactive program for CKM application is shown in \mbox{Fig.~\ref{program_cons_nlos}} and elaborated as follows:
\begin{itemize}
    \item[$T_0$:] ISAC BS periodically conducts environment sensing with the sensing frames and performs initial access to the UE within the experiment area to obtain its location. 
    \item[$T_1$:] ISAC BS continuously tracks the dynamic scatterers and the UE.
    When UE in the LoS area is temporarily obstructed by a dynamic scatterer, the movement path of UE is predictable. 
    Therefore, in this case, the ISAC BS can directly align the beam according to the CKM without the location feedback from the UE.
    When UE enters the NLoS area, the ISAC BS is unable to actively sense its location and requires frequent beam training for maintaining a connection with the UE. 
    In this case, the UE proactively reports its location, enabling the BS to directly acquire the optimal beam index corresponding to the UE's location from a pre-constructed CKM, thereby obviating the need for beam training.
    \item[$T_2$:] When UE is in the LoS area, the ISAC BS utilizes location-based beam alignment.
    And then, the ISAC BS transmits data to UE while continuously tracking its location by receiving the echo.
\end{itemize}

\begin{figure*}[!ht]
	\centering
        \subfigure[The $1st$ strongest beam indices.]{\label{exp_ckm1}
        \includegraphics[width=0.31\textwidth]{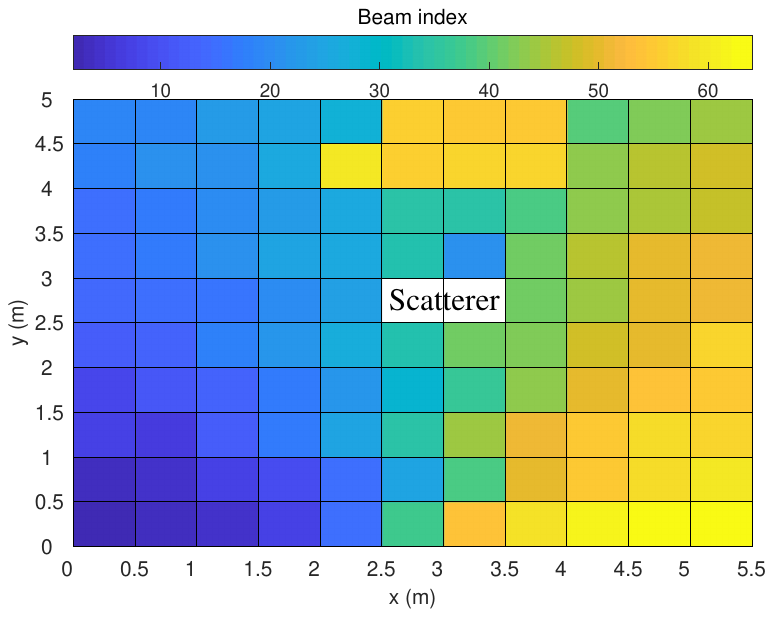}
		}
	\hfil
	\subfigure[The $2nd$ strongest beam indices.]{\label{exp_ckm2}
        \includegraphics[width=0.31\textwidth]{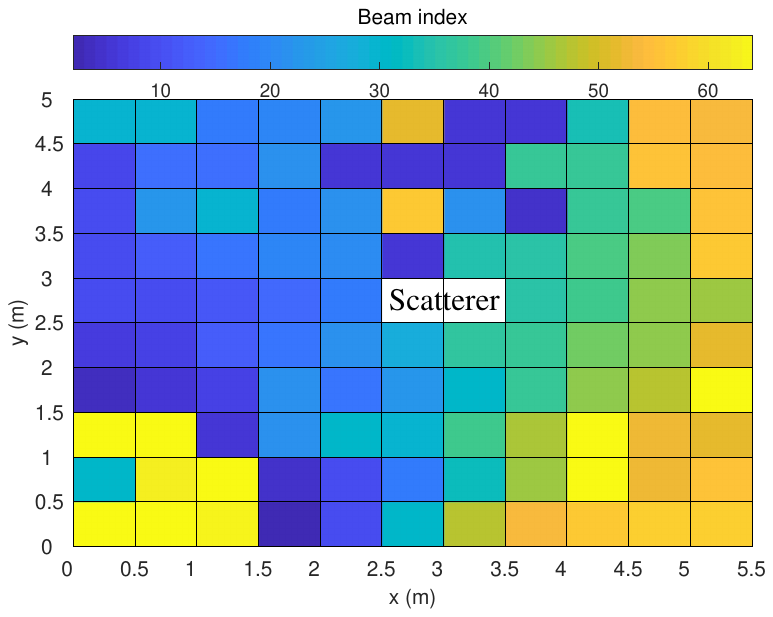}
        } 
        \hfil
        \subfigure[The $3rd$ strongest beam indices.]{\label{exp_ckm3}
        \includegraphics[width=0.31\textwidth]{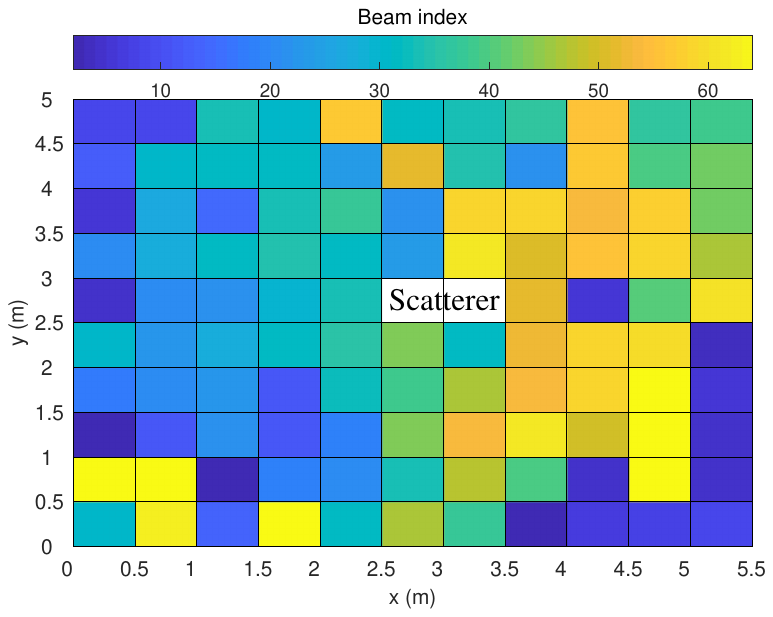}
        }
    \caption{Experimental result for ISAC-based BIM construction.}
    \label{exp_ckm}
\end{figure*}

In particular, although beam alignment based on exhaustive search is the most common strategy, we did not compare its performance in our experiments. 
The reason is twofold. 
From the perspective of RSS measurement, the exhaustive search strategy performs exactly the same as CKM.
On the other hand, from the perspective of resource consumption, the performance of exhaustive search strategy highly depends on waveform design. 
In mmWave ISAC scenarios, improving the frequency of exhaustive search will sacrifice communication performance. 
Comparing the performance of exhaustive search strategy may introduce many variables and make experimental design more complex. 
Therefore, we plan to compare the performance of multiple beam alignment strategies in terms of switching speed, resource consumption, and other aspects in future experiments.

\section{Experimental results}\label{ss_exp_res}
\subsection{BIM construction}
The experimental results of the BIM construction are shown in \mbox{Fig.~\ref{exp_ckm}}. 
UE reported the three strongest beam indices for all grids within the experiment area and then the ISAC BS stored them into the server. 
The selection of grid size depends on the resolution of the ISAC system. 
For a mmWave phased array with a beamwidth of $\SI{5}{\degree}$ used in this experiment, the grid size is set to $\qtyproduct{0.5 x 0.5}{\metre}$.
The color of each grid in \mbox{Fig.~\ref{exp_ckm}} represents the beam index, with darker colors indicating smaller beam indices and lighter colors indicating larger beam indices.  
Blank grids in the figure represent areas where mapping is not possible due to obstacles present in those grids.

It can be observed from the LoS area with y-coordinates of $\SIrange{0}{3.2}{\meter}$ in \mbox{Fig.~\ref{exp_ckm1}} that the strongest beam index increases from left to right.
The beam index of UE is more leftward when it is located in a grid of BIM with darker color, while it is more rightward in a grid with lighter color. 
Due to the use of DFT codebook, the variation of beam index is most pronounced in the direction of about $\SI{90}{\degree}$.
In addition, there are some conspicuous grids with beam indices of around $55$ in the NLoS area behind the scatterer, indicating that the strongest path within this area is provided by the metal reflector on the right side.

The second strongest BIM, as shown in \mbox{Fig.~\ref{exp_ckm2}}, exhibits a similar variation trend to the strongest BIM within the LoS area of approximately $\SIrange{45}{135}{\degree}$ due to the presence of scatterer directly ahead. 
However, in the wider angle range closer to $\SI{0}{\degree}$ and $\SI{180}{\degree}$, dominant beams originate from other directions. 
Smaller beam indices appear in the NLoS area behind the scatterer, indicating that the second strongest beam mainly originates from environmental reflections on the left hand side.

The third strongest BIM, as shown in \mbox{Fig.~\ref{exp_ckm3}}, reveals some environmental characteristics. 
For example, when UE is located within the x-coordinate range of $\SIrange{3.5}{5}{\meter}$, the beam indices along y-axis is close, indicating that there is likely a huge wall outside the experiment area on the right side, which is consistent with the actual environment.

\subsection{CGM construction and ISAC sensing results}
\begin{figure}[htbp]
\centering
	{\includegraphics[width=0.36\textwidth]{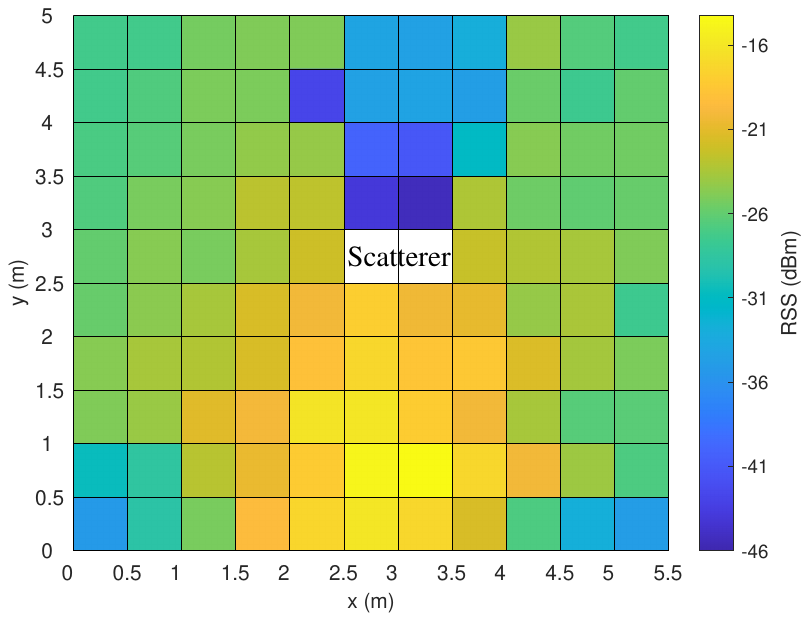}}
	\caption{Experimental result for ISAC-based CGM construction.}
	\label{exp_cgm}
\end{figure}

By applying the BIM on the ISAC BS, the CGM obtained from UE measurements is shown as \mbox{Fig.~\ref{exp_cgm}}.
At a fixed angle, the RSS decreases as the distance between UE and BS increases. 
At a fixed distance, the RSS decreases as the angle expands from $\SI{90}{\degree}$ to both sides due to poorer focusing ability of DFT codebook on both sides. 
The two blank grids at the center of the figure represent the scatterer, and the areas behind it are all NLoS.
In the NLoS area, the four adjacent grids to the scatterer experience significant fading, followed by illumination of subsequent grids. 
This is due to the fact that the illuminated grids fall within the coverage area of a metal reflector on the right side.

\begin{figure}[htbp]
\centering
	{\includegraphics[width=0.36\textwidth]{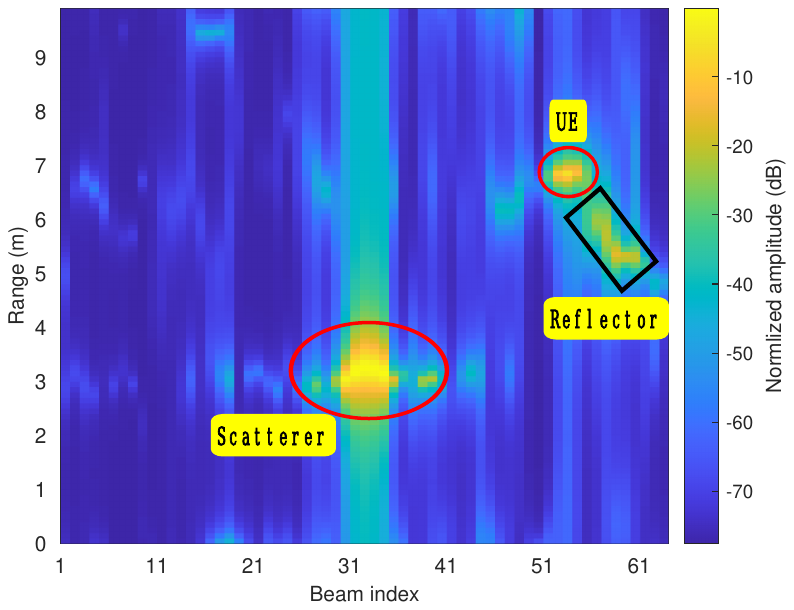}}
	\caption{Experimental result for ISAC sensing.}
	\label{exp_radar}
\end{figure}

ISAC BS uses sensing frames to sense the environment, and employs one-dimensional Capon beamforming algorithm to process the echo of each subframe. 
The sensing results from 64 subframes are then concatenated to obtain the range-angle spectrum shown in \mbox{Fig.~\ref{exp_radar}}.
In which three main targets can be observed: a scatterer in front of the ISAC BS, a metal reflector on the right side, and an UE is stationary at $(6.4,5.6)$. 
As the scatterer located at $(3.2,3.2)$ is closest to the BS, it presents the most significant reflected signal strength in the sensing result.
The UE at $(6.4,5.6)$ appears as a smaller, highlighted target. Although its RCS is relatively small, the large amount of metal structure on it provides an ideal performance on mmWave reflection.
From the sensing results, the metal reflector located at $(7.0,4.0)$ does not have a distinct radar pattern as we expect. 
In order to provide a reflection path for the UE in the NLoS area, the metal reflector is placed at an angle that is not directly toward BS. 
Due to the weak scattering properties of the smooth copper surface, most of the energy from the incident mmWave beam cannot be reflected along the original path, and only a small portion is reflected by the irregular frame of the metal reflector, resulting in a inconspicuous pattern in the range-angle spectrum.

\subsection{Performance comparison}
\begin{figure}[htbp]
\centering
	{\includegraphics[width=0.36\textwidth]{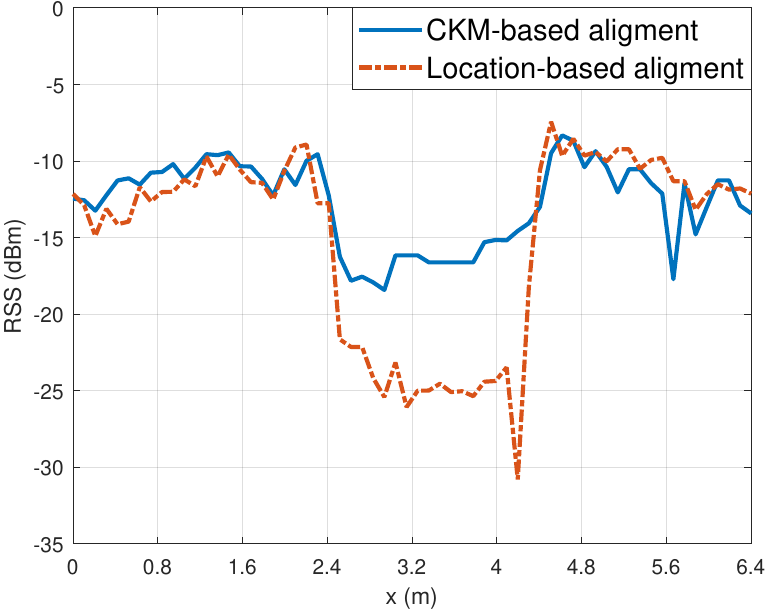}}
	\caption{Performance comparison of CKM-based and location-based beam alignment.}
	\label{fig_exp_beam}
\end{figure}

Finally, we compare the performance of CKM-based beam alignment with the baseline of location-based beam alignment. 
As shown in \mbox{Fig.~\ref{fig_exp_beam}}, within the NLoS area at x-coordinates of $\SIrange{2.4}{4.5}{\meter}$, location-based beam alignment is blocked and experiences an attenuation of approximately $\SI{15}{dBm}$ in RSS. 
On the other hand, CKM-based beam alignment, due to its exploitation of environmental prior information, switches the beam to the strongest path resulting in only about $\SI{5}{dB}$ attenuation in RSS.
However, at some locations within the LoS area, the performance of CKM-based beam alignment is worse than the baseline. 
This is mainly because that the BIM construction is grid-based, and in this experiment, the grid resolution is \SI{0.5}{\meter}. 
Therefore, in some cases, the actual location of the UE corresponds to a different optimal beam index than the one recorded in the BIM, which leads to lower received power.
The experimental results indicate that the CKM can effectively handle environmental changes with ISAC and improve communication and sensing performance through pre-learned channel knowledge.

\section{Conclusion}\label{ss_conclu}

In this paper, we elaborate the principle of the ISAC-based CKM system, and propose a specific ISAC waveform suitable for a single-RF chain mmWave phased array. 
Based on OFDM waveforms, three mainstream ISAC signal processing algorithms are introduced, and two types of CKM (BIM and CGM) are constructed based on the estimated parameters. 
Finally, a prototype system of ISAC-based CKM is built, and its performance is tested in various scenarios. 
Through the experiments, this paper reveals the inherently mutualistic relationship between ISAC and CKM, laying the foundation for their applications in future wireless communication systems.





\bibliographystyle{IEEEtran}
\bibliography{IEEEabrv,reference}

%

\begin{IEEEbiography}[{\includegraphics[width=1in,height=1.25in,clip,keepaspectratio]{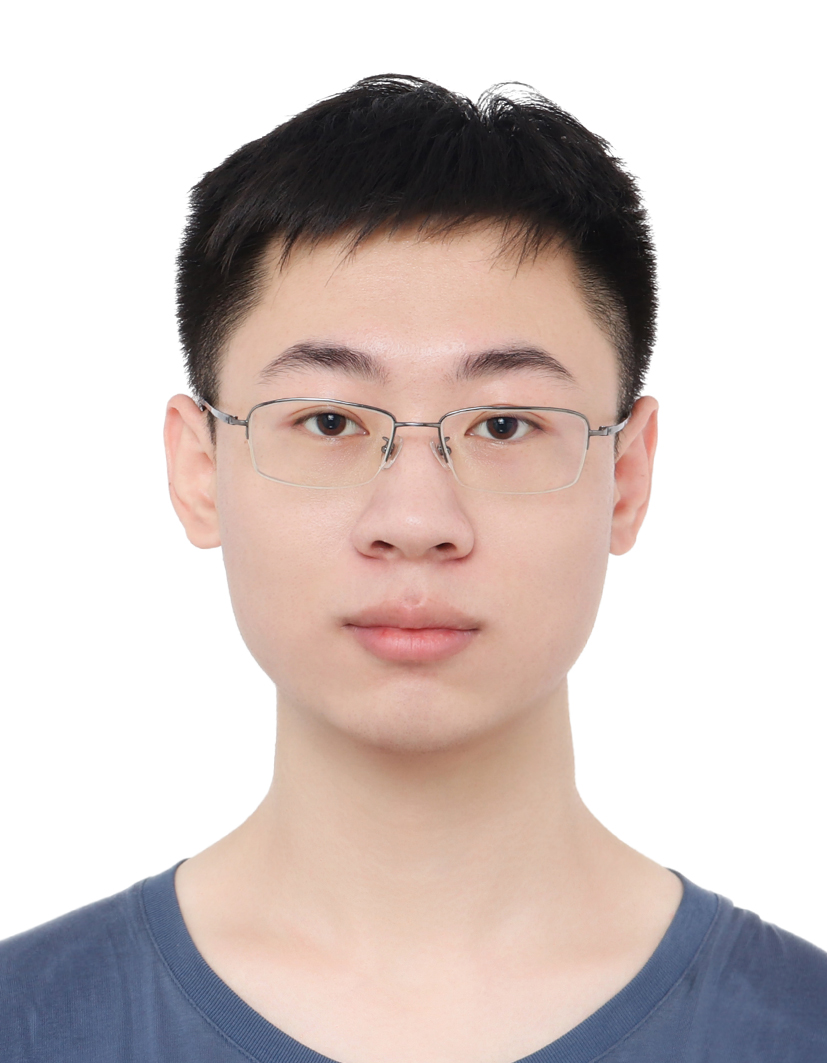}}]{Chaoyue Zhang} received the B.S. degree in electronic science and technology from Nanjing University of Posts and Telecommunications, China, in 2021, and the M.S. degree in electronic and communication engineering from Southeast University, China, in 2024. 
He is currently pursuing the Ph.D. degree with the National Mobile Communications Research Laboratory, Southeast University, Nanjing. 
His research interests include integrated sensing and communication (ISAC).
\end{IEEEbiography}

\begin{IEEEbiography}[{\includegraphics[width=1in,height=1.25in,clip,keepaspectratio]{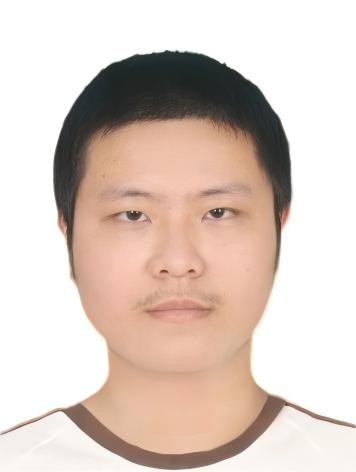}}]{Zhiwen Zhou} received the B.S. degree in electronic and information engineering from Nanjing University of Posts and Telecommunications, China, in 2022. He is currently pursuing the M.S. degree with the National Mobile Communications Research Laboratory, Southeast University, Nanjing. His research interests include integrated sensing and communication (ISAC) and extremely large-scale arrays (XL-arrays).
\end{IEEEbiography}

\begin{IEEEbiography}[{\includegraphics[width=1in,height=1.25in,clip,keepaspectratio]{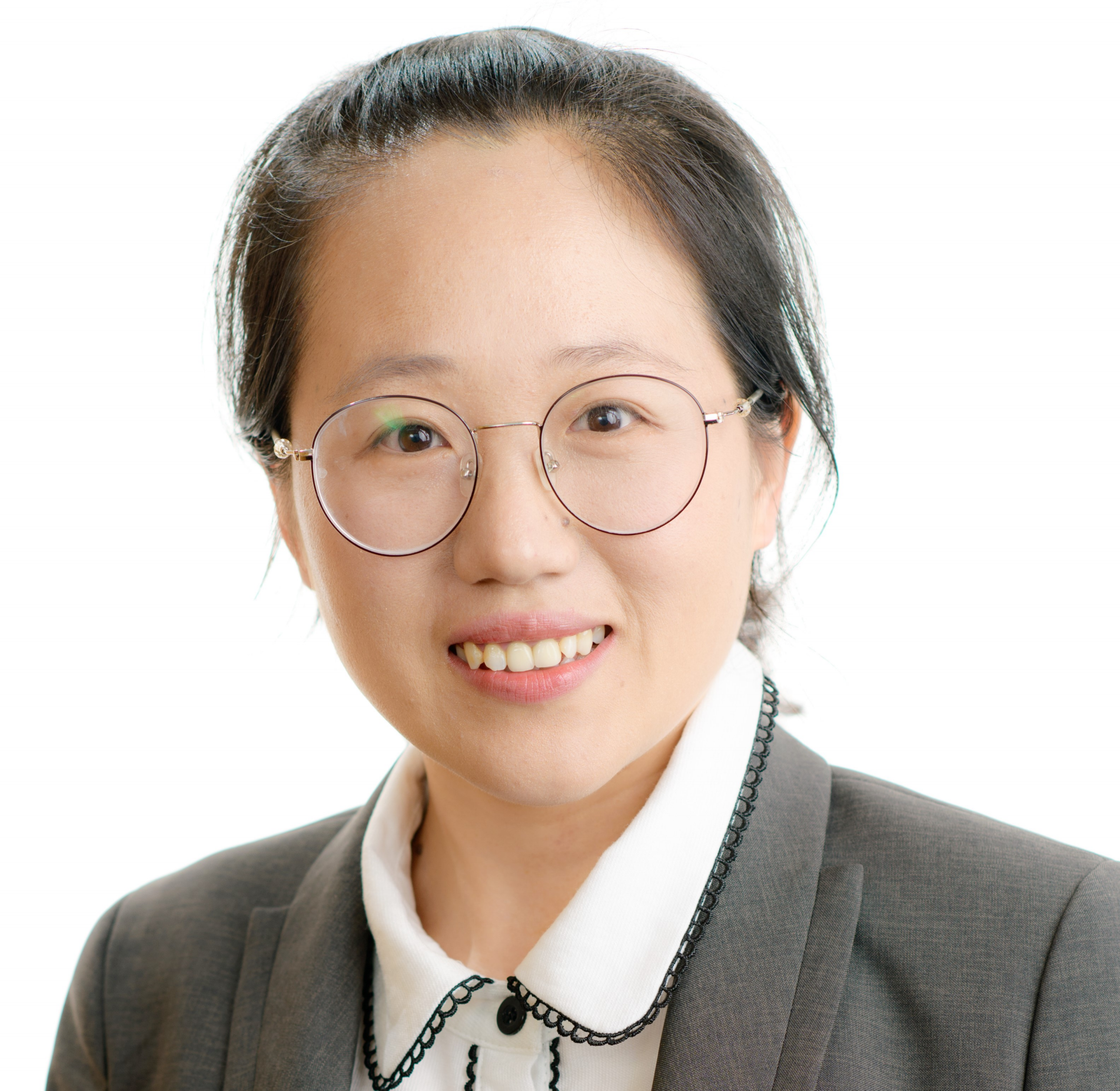}}]{Xiaoli Xu} is with the School of Information Science and Engineering, Southeast University, China. She received the Bachelor of Engineering (First-Class Honours) and Ph.D. degrees from Nanyang Technological University, Singapore, in 2009 and 2015, respectively. From 2015 to 2018, she was a Research Fellow at the Nanyang Technological University. From 2018 to 2019, she was a Postdoctoral Research Associate at the University of Sydney, Australia. Her research interests include network coding, low-latency wireless communications, semantic communications and integrated sensing and communication. 
\end{IEEEbiography}

\begin{IEEEbiography}[{\includegraphics[width=1in,height=1.25in,clip,keepaspectratio]{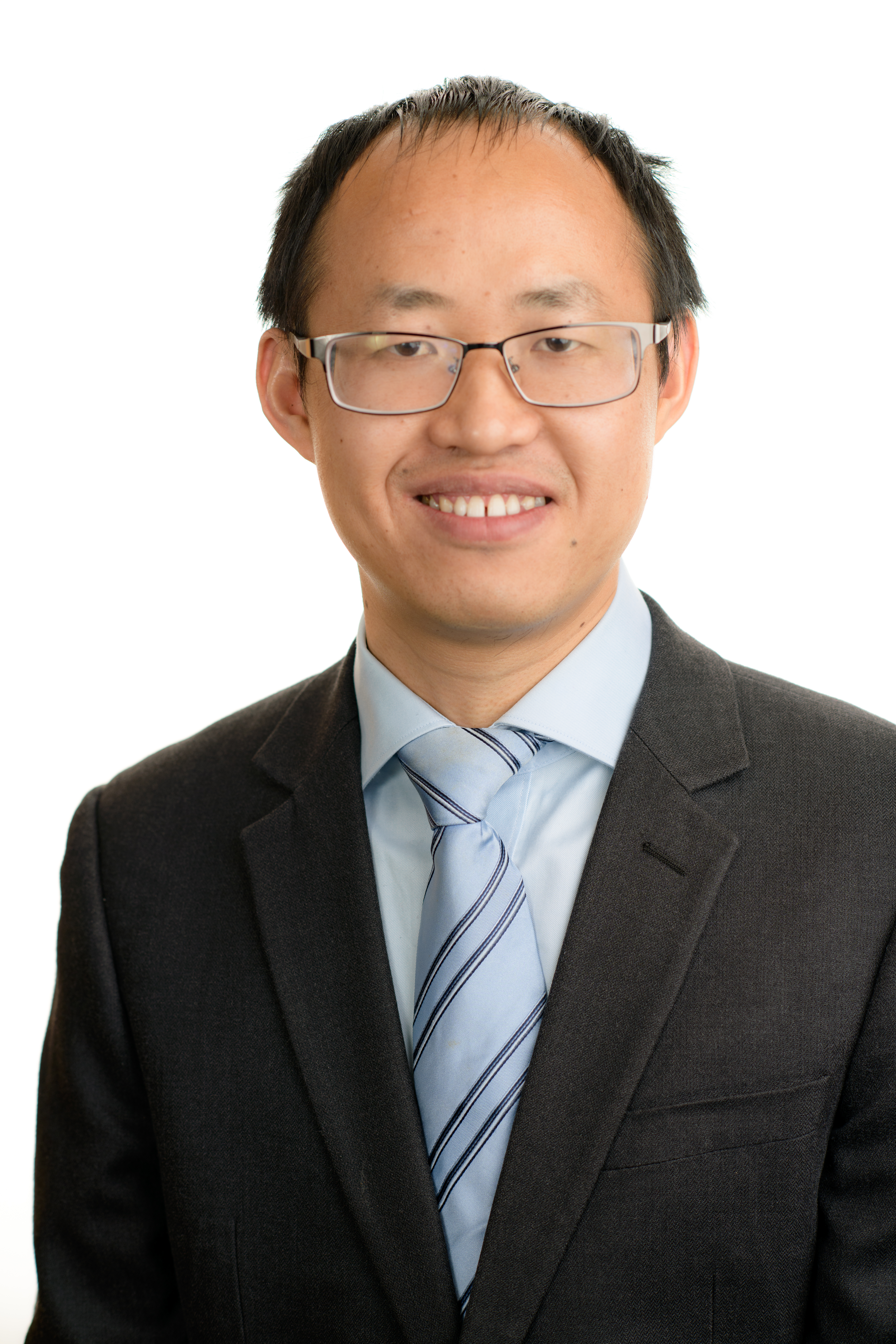}}]{Yong Zeng} (S’12-M’14-SM’22-F’25) is a Chief Young Professor with the National Mobile Communications Research Laboratory, Southeast University, China, and also with the Purple Mountain Laboratories, Nanjing, China. He received the Bachelor of Engineering (First-Class Honours) and Ph.D. degrees from Nanyang Technological University, Singapore. From 2013 to 2018, he was a Research Fellow and Senior Research Fellow at the Department of Electrical and Computer Engineering, National University of Singapore. From 2018 to 2019, he was a Lecturer at the School of Electrical and Information Engineering, the University of Sydney, Australia.\\
Prof. Zeng was listed as Highly Cited Researcher by Clarivate Analytics for six consecutive years (2019-2024). He is the recipient of the Australia Research Council (ARC) Discovery Early Career Researcher Award (DECRA), 2020 \& 2024 IEEE Marconi Prize Paper Award in Wireless Communications, 2018 IEEE Communications Society Asia-Pacific Outstanding Young Researcher Award, 2020 \& 2017 IEEE Communications Society Heinrich Hertz Prize Paper Award, 2021 IEEE ICC Best Paper Award, and 2021 China Communications Best Paper Award. He serves as an Editor for IEEE Transactions on Communications, IEEE Transactions on Mobile Computing, IEEE Communications Letters and IEEE Open Journal of Vehicular Technology, Leading Guest Editor for IEEE Wireless Communications on ``Integrating UAVs into 5G and Beyond'' and China Communications on ``Network-Connected UAV Communications''. He is the Symposium Chair for IEEE Globecom 2021 Track on Aerial Communications, the workshop co-chair for ICC 2018-2023 workshop on UAV communications, the tutorial speaker for Globecom 2018/2019 and ICC 2019 tutorials on UAV communications. Prof. Zeng proposed the concept of channel knowledge map (CKM) and the transmission method of delay-Doppler alignment modulation (DDAM). He has published more than 200 papers, which have been cited by more than 30,000 times based on Google Scholar. Prof. Zeng was elevated to IEEE Fellow “for contributions to unmanned aerial vehicle communications and wireless power transfer”.
\end{IEEEbiography}

\begin{IEEEbiography}[{\includegraphics[width=1in,height=1.25in,clip,keepaspectratio]{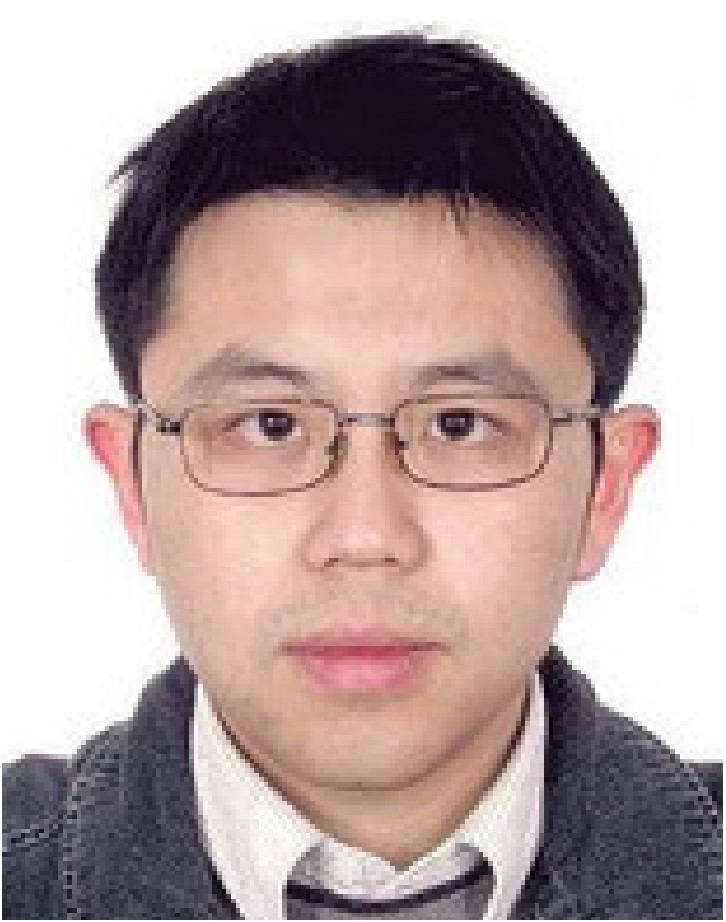}}]{Zaichen Zhang} (Senior Member, IEEE) was born in Nanjing, China, in 1975. He received the B.S. and M.S. degrees in electrical and information engineering from Southeast University, Nanjing, China, in 1996 and 1999, respectively, and the Ph.D. degree in electrical and electronic engineering from The University of Hong Kong, Hong Kong, China, in 2002. From 2002 to 2004, he was a Post-Doctoral Fellow with the National Mobile Communications Research Laboratory, Southeast University. He joined the School of Information Science and Engineering, Southeast University, in 2004, where he is currently a Professor. He has published more than 250 articles and issued more than 70 patents. His current research interests include 6G mobile communication systems, optical mobile communications, and quantum information technologies.
\end{IEEEbiography}

\begin{IEEEbiography}[{\includegraphics[width=1in,height=1.25in,clip,keepaspectratio]{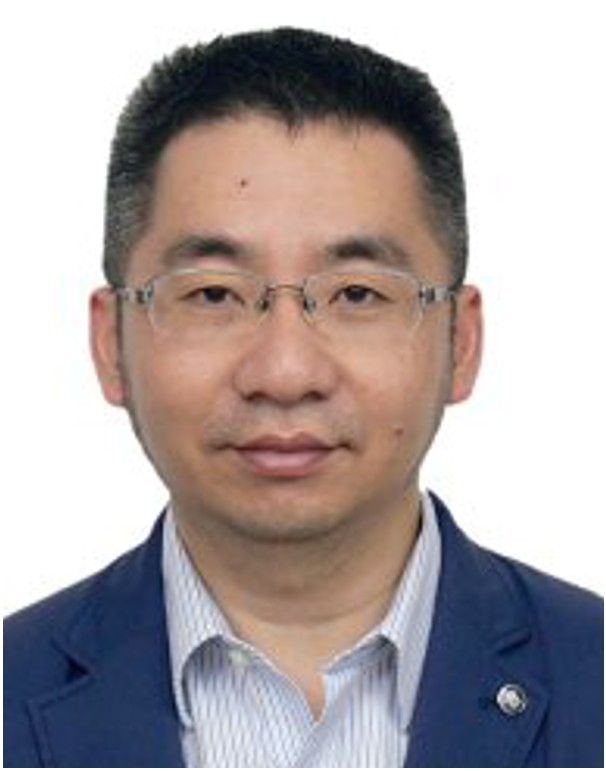}}]{Shi Jin} (Fellow, IEEE) received the Ph.D. degree in communications and information systems from Southeast University in 2007. From June 2007 to October 2009, he was a Research Fellow with University College London, U.K., Adastral Park Research Campus. He is currently a Faculty Member of the National Mobile Communications Research Laboratory, Southeast University. His research interests include wireless communications, random matrix theory, and information theory. He and his co-authors have been awarded the 2011 IEEE Communications Society Stephen O. Rice Prize Paper ward in the field of communication theory, the IEEE Vehicular Technology Society 2023 Jack Neubauer Memorial Award, and the 2022 Best Paper Award and the 2010 Young Author Best Paper Award by the IEEE Signal Processing Society. He serves as an Area Editor for IEEE TRANSACTIONS ON COMMUNICATIONS and \textit{IET Electronics Letters}. Previously, he was an ssociate Editor of IEEE TRANSACTIONS ON WIRELESS COMMUNICATIONS, IEEE COMMUNICATIONS LETTERS, and \textit{IET Communications}.
\end{IEEEbiography}

\vfill
\end{document}